%% file: 0_main.tex
\documentclass[sigconf]{acmart}
\renewcommand\footnotetextcopyrightpermission[1]{} % removes footnote with conference information in first column
\usepackage{amsfonts,amsmath}
\usepackage{algorithm}% http://ctan.org/pkg/algorithms
\usepackage[noend]{algpseudocode}% http://ctan.org/pkg/algorithmicx
\usepackage{bm}
\usepackage{booktabs}
\usepackage{soul}
\usepackage{multirow}
\usepackage{subfig}
\usepackage[flushleft]{threeparttable}
\usepackage{adjustbox}
\usepackage{fancyhdr}
\usepackage{xcolor}
\usepackage[flushleft]{threeparttable}
\usepackage{tikz}
\usepackage{outlines}
\usepackage{balance}
\usepackage{tcolorbox}
\usepackage{wrapfig}
\usepackage{blindtext}
\usepackage{xcolor}
\usepackage{balance}
\AtBeginDocument{%
  \providecommand\BibTeX{{%
    \normalfont B\kern-0.5em{\scshape i\kern-0.25em b}\kern-0.8em\TeX}}}

\begin{document}

%%
%% The "title" command has an optional parameter,
%% allowing the author to define a "short title" to be used in page headers.
\title{Low-latency Mini-batch GNN Inference on \\ CPU-FPGA Heterogeneous Platform}

%%
%% The "author" command and its associated commands are used to define
%% the authors and their affiliations.
%% Of note is the shared affiliation of the first two authors, and the
%% "authornote" and "authornotemark" commands
%% used to denote shared contribution to the research.
\author{Bingyi Zhang}
\email{bingyizh@usc.edu}
\affiliation{%
  \institution{University of Southern California}
  \city{Los Angeles}
  \state{California}
  \country{USA}
}

\author{Hanqing Zeng}
\email{zengh@usc.edu}
\affiliation{%
  \institution{University of Southern California}
  \city{Los Angeles}
  \state{California}
  \country{USA}
}

\author{Viktor Prasanna}
\email{prasanna@usc.edu}
\affiliation{%
  \institution{University of Southern California}
  \city{Los Angeles}
  \state{California}
  \country{USA}
}
%%
%% By default, the full list of authors will be used in the page
%% headers. Often, this list is too long, and will overlap
%% other information printed in the page headers. This command allows
%% the author to define a more concise list
%% of authors' names for this purpose.
\renewcommand{\shortauthors}{Trovato and Tobin, et al.}

%%
%% The abstract is a short summary of the work to be presented in the
%% article.
\begin{abstract}
  Mini-batch inference of Graph Neural Networks (GNNs) is a key problem in many real-world applications. Recently, a GNN design principle of model depth-receptive field decoupling has been proposed to address the well-known issue of neighborhood explosion.  Decoupled GNN models achieve higher accuracy than original models and demonstrate excellent scalability for mini-batch inference.
 
  We map Decoupled GNNs onto CPU-FPGA heterogeneous platforms to achieve low-latency mini-batch inference. On the FPGA platform, we design a novel  GNN hardware accelerator with  an adaptive datapath denoted Adaptive Computation Kernel (ACK) that can execute various computation kernels of GNNs with low-latency: (1) for dense computation kernels expressed as matrix multiplication,  ACK works as a systolic array with fully localized connections, (2) for sparse computation kernels,   ACK follows the scatter-gather paradigm and works as multiple parallel pipelines to support the irregular connectivity of graphs.  The proposed task scheduling hides the CPU-FPGA data communication overhead to reduce the inference latency. We develop a fast design space exploration algorithm to generate a single accelerator for multiple target GNN models.   We  implement our accelerator on a state-of-the-art CPU-FPGA platform and evaluate the performance using three representative models (GCN, GraphSAGE, and GAT).  Results show that our CPU-FPGA  implementation achieves  $21.4-50.8\times$, $2.9-21.6\times$, $4.7\times$ latency reduction compared with state-of-the-art implementations on CPU-only, CPU-GPU and CPU-FPGA platforms.

\end{abstract}

%%
%% The code below is generated by the tool at http://dl.acm.org/ccs.cfm.
%% Please copy and paste the code instead of the example below.
%%
% \begin{CCSXML}
% <ccs2012>
%   <concept>
%       <concept_id>10010520.10010521.10010528</concept_id>
%       <concept_desc>Computer systems organization~Parallel architectures</concept_desc>
%       <concept_significance>500</concept_significance>
%       </concept>
%  </ccs2012>
% \end{CCSXML}

% \ccsdesc[500]{Computer systems organization~Parallel architectures}

%%
%% Keywords. The author(s) should pick words that accurately describe
%% the work being presented. Separate the keywords with commas.

% \keywords{Graph neural network, Mini-batch inference, FPGA}

%% A "teaser" image appears between the author and affiliation
%% information and the body of the document, and typically spans the
%% page.

%%
%% This command processes the author and affiliation and title
%% information and builds the first part of the formatted document.
\maketitle

\input{1_introduction}

\input{2_background}

\input{3_Design_Methodology}
\input{4_Hardware_Architecture}

\input{5_Optimizations}
\input{6_Experiments}
\input{7_Discussion}
\input{8_Conclusion}

\bibliographystyle{acmart/ACM-Reference-Format}
\bibliography{citation} 

%%
%% If your work has an appendix, this is the place to put it.

\end{document}

%% file: 1_introduction.tex
\section{Introduction}
\label{sec:intro}

Graph Neural Networks (GNNs) have become a revolutionary technique in graph-based machine learning. GNNs  outperform traditional algorithms in various applications \cite{hu2020open}. Recently, many vendors have adopted GNNs in their commercial  systems, such as  recommendation systems \cite{yang2019aligraph}, social media \cite{lerer2019pytorch}, knowledge databases \cite{mernyei2020wiki}, etc. In these systems, data are represented as graphs, where vertices correspond to entities and edges encode the relationship among entities. A fundamental task is mini-batch GNN inference: \emph{given a set of target vertices, infer the embeddings (vector representation) of the vertices with low-latency.}

 \begin{figure}[h]
    \centering
    \includegraphics[width=7.5cm]{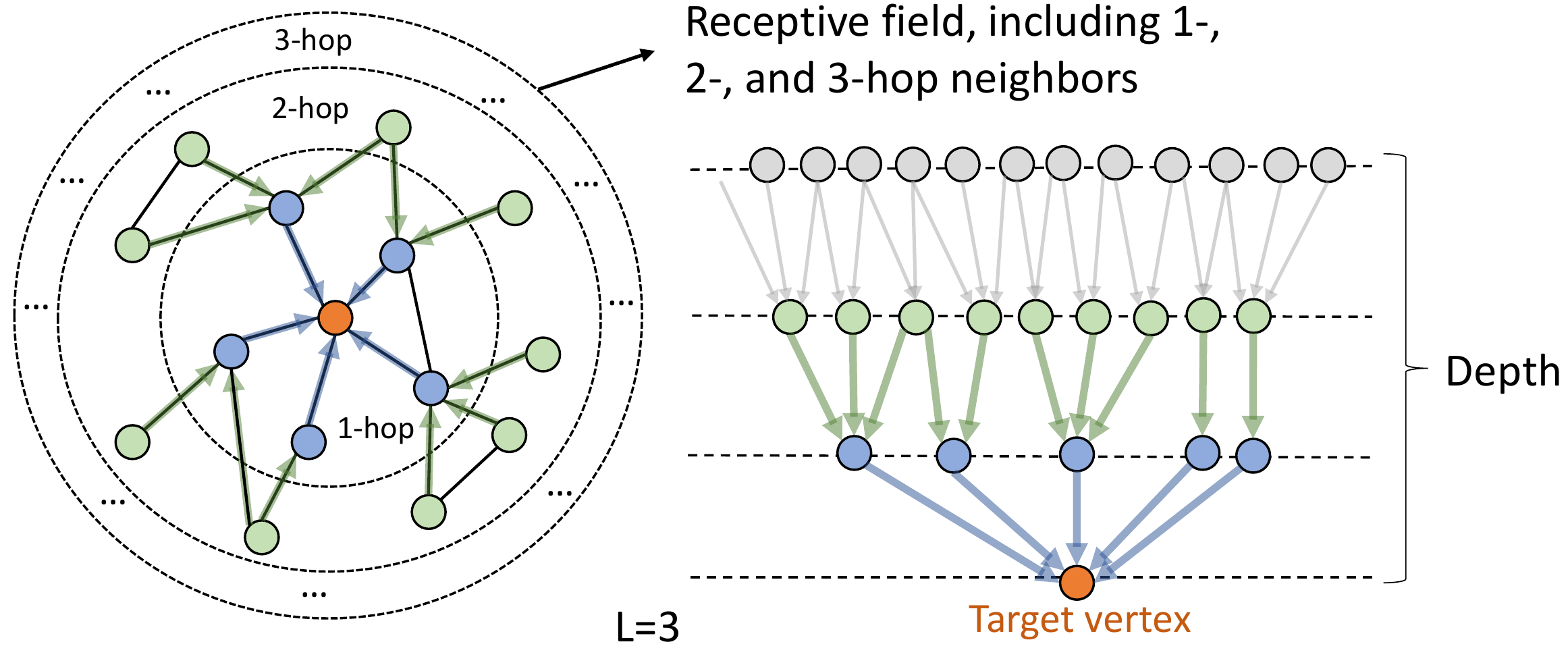}
    \vspace{-0.3cm}
    \caption{ Recursive message passing of GNNs results in exponential computation and communication cost, and low computation-to-communication (C2C) ratio}
    \vspace{-0.5cm}
     \label{fig:receptive-field}
\end{figure}

There are two major challenges for mini-batch inference. \textbf{(1) neighborhood explosion}: 
% a well-known issue in mini-batch GNN inference is \emph{neighborhood explosion}.
the widely used GNNs, such as GCN \cite{kipf2016semi}, GraphSAGE \cite{hamilton2017inductive}, GIN \cite{xu2018powerful},  GAT \cite{velivckovic2017graph}, follow the message-passing paradigm that in a $L$-layer model, a vertex recursively aggregates information from its $L$-hop neighbors.  The receptive field is defined as the set of neighbors passing messages to the target vertex. In the above example (Figure \ref{fig:receptive-field}), the receptive field consists of all the vertices within $L$-hop. In a large graph, the size of the receptive field quickly explodes w.r.t. model depth $L$. Therefore,  mini-batch GNN inference suffers from two issues.  First, the computation and communication costs grow exponentially with the depth of GNN. This hinders the deployment of deeper GNNs on memory constrained accelerators. It has been proven \cite{gallicchio2020fast, li2021deepgcns, zeng2021decoupling} that deeper GNNs have higher accuracy than shallower 
ones. Second, the computation-to-communication (C2C) ratio is low, thus making it not suitable for hardware acceleration.  \textbf{(2) load imbalance}: GNN computation involves various kinds of kernels, including dense computation kernels and sparse computation kernels. To support various kernels, previous work \cite{yan2020hygcn, zeng2020graphact, zhang2021boostgcn}  designed \emph{hybrid} accelerators that for each kernel, a dedicated  hardware module is designed and initialized independently. For example, in  GraphACT \cite{zeng2020graphact}, the Feature Aggregation Module executes the sparse kernel and Feature Transformation Module executes 
the dense kernel. However, it is challenging to achieve
load balance among hardware modules in a hybrid accelerator. For example, the workload of feature aggregation is unpredictable and depends on the connectivity of the input graphs. The load imbalance leads to hardware under-utilization and extra latency.

 \begin{figure}[h]
    \centering
    \includegraphics[width=8cm]{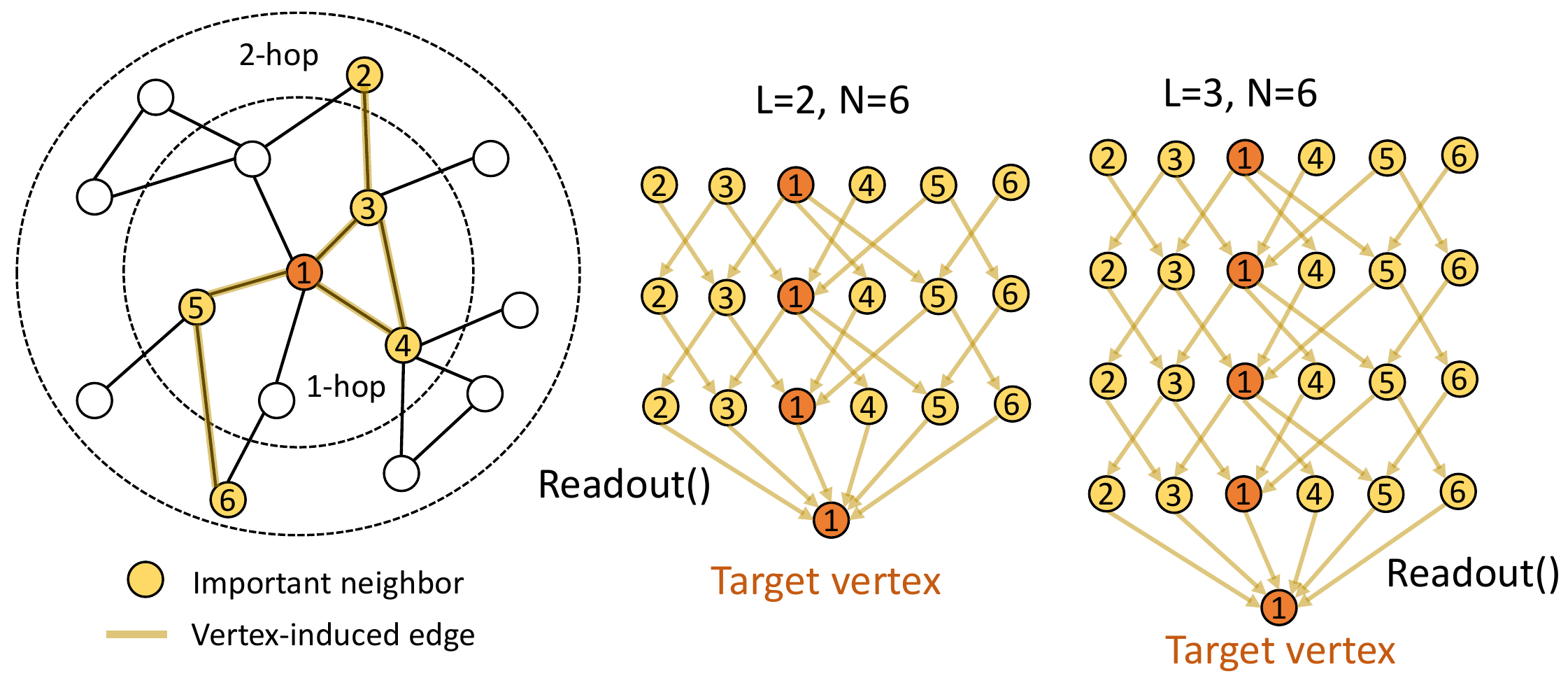}
    \vspace{-0.3cm}
    \caption{ An example of Decoupled GNN model}
    \vspace{-0.3cm}
     \label{fig:decoupled model}
\end{figure}

Recently, \emph{model depth-receptive field decoupling} \cite{zeng2021decoupling} is proposed to resolve the \emph{neighborhood explosion} challenge. Under the decoupling principle,  GNN depth is independent of receptive field. As shown in Figure 2, for a target vertex, it selects a small number of important neighbors as the receptive field $\mathcal{N}$, and then applies an arbitrary deep GNN upon such pre-defined $\mathcal{N}$. The key property of a Decoupled GNN is that the size of $\mathcal{N}$ remains fixed while the GNN becomes deeper, thus reducing the computation complexity from exponential to linear w.r.t. model depth. Each layer only propagate information within $\mathcal{N}$. Compared with original GNN models (Figure \ref{fig:receptive-field}), the Decoupled GNN models theoretically lead to significantly less computation cost and memory bandwidth requirement and thus are well-suited for hardware acceleration.

While there have been many GNN accelerators \cite{yan2020hygcn, geng2020awb, zhang2021boostgcn, liang2020deepburning, zeng2020graphact, lin2021hp} proposed, none of them is designed or optimized for low-latency mini-batch inference.  
Specifically, \cite{yan2020hygcn, geng2020awb, zhang2021boostgcn, liang2020deepburning} are for \emph{full-graph} inference and \cite{zeng2020graphact} is for mini-batch \emph{training}. 
Full-graph GNN inference has very different computation characteristics from mini-batch inference. In full-graph inference, all vertices in the graph are target vertices. Through careful vertex reordering \cite{zhang2020hardware, geng2021gcn} and graph partitioning \cite{yan2020hygcn, zhang2021boostgcn},  full-graph execution can achieve high data reuse by exploiting common neighbors. However, in mini-batch inference, it is more challenging to improve data reuse since the target vertices rarely share common neighbors. 
% it is highly possible that the target vertices do not share common neighbors, thus resulting in low data reuse. 
In addition, GraphACT \cite{zeng2020graphact} is built on specific training algorithm \cite{zeng2019graphsaint} to improve computation-to-communication ratio \emph{only} during training. 
While the computation pipeline of GraphACT can be adapted to inference, its performance may be sub-optimal due to load imbalance (challenge (2) above).

% FPGA is a promising hardware platform for accelerating many real-world applications \cite{danopoulos2021utilizing,koliogeorgi2021fpga,haghi2021fpga} due to its fine-grained data parallelism. 
%  While the \emph{neighbor explosion} issue can be resolved through GNN model transformation, it is still inefficient to execute mini-batch GNN inference 
%   on general purpose processor (GPPs). Because  the datapath and memory organization of GPPs cannot efficiently execute various computation kernels (e.g., feature aggregation, feature transformation, attention) in GNN models (e.g., GAT, GraphSAGE, GCN). 
In this paper, we map the inference process of Decoupled models on CPU-FPGA  platforms  for low-latency mini-batch inference. To this end, we design a unified hardware accelerator consisting of (1) adaptive datapath that can execute various computation kernels of GNNs with low-latency, thus overcoming the load-imbalance challenge, (2) memory organization that can hide data communication overhead to further reduce the inference latency. Our main contributions are: 
\begin{itemize}
    \item By analyzing the computation and communication cost of mini-batch GNN inference, we identify ``model depth-receptive field'' decoupling as a key model design technique towards low-latency accelerator design.
    \item We propose a system design on CPU-FPGA platforms to achieve low-latency mini-batch inference:
     \begin{itemize}
        \item We develop a novel hardware accelerator with two execution modes  that can execute various GNN computation kernels (e.g., feature aggregation, feature transformation, attention) with low-latency.
        \item We customize the memory organization to achieve low-latency data communication, and proposed double/triple buffering techniques to hide communication overhead.
        \item We perform task scheduling to hide the CPU-FPGA data movement overhead.
        % \item On the CPU platform, to further reduce inference latency, we perform task scheduling to pipeline the important neighbor identification on host processor and the GNN computations on FPGA accelerator. 
     \end{itemize}
    \item We develop a design space exploration algorithm %(Section \ref{subsec:design-space-exploration}) 
    that given 1) specification of the target FPGA device, and 2) a set of target GNN models with various depths and receptive field sizes, it generates a \emph{single} hardware that achieves low-latency inference without reconfiguration.
    \item  We implement our hardware accelerator on a state-of-the-art CPU-FPGA platform and evaluate the performance using three representative models (GCN, GraphSAGE, GAT). Experiments show that our CPU-FPGA implementation achieves $21.4-50.8\times$, $2.9-21.6\times$, $4.7\times$ speedup compared with state-of-the-art implementations on CPU-only, CPU-GPU and CPU-FPGA platforms.
\end{itemize}

%% file: 2_background.tex
\section{background}

\subsection{Field Programmable Gate Array}
Recently, Field Programmable Gate Array (FPGA) has been extensively studied for accelerating machine learning tasks \cite{geng2020awb, zhang2021boostgcn, zeng2020graphact}. A FPGA device deployed in a cloud has significant hardware resources, including Lookup Tables (LUTs), Digital Signal Processing units (DSPs), on-chip memories (BRAMs, URAMs) and programmable interconnections. 
% FPGA provides users with the opportunities to build the highly parallelized hardware architecture through utilizing the large number of DSPs and customizing the on-chip memory organization. 
Programmability of FPGA allows users to exploit the fine-grained data parallelism in a computation task.  An FPGA is more attractive for low-latency computation compared with GPU which has coarse-grained thread-level parallelism. 

\noindent \textbf{GNN acceleration on FPGA}: As discussed in  Section \ref{sec:intro}, previous GNN accelerators on FPGA such as GraphACT \cite{zeng2020graphact}, BoostGCN \cite{zhang2021boostgcn}, Deepburning-GL \cite{liang2020deepburning} , AWB-GCN \cite{lin2021hp}, are not suitable for mini-batch GNN inference. 
In this work, we develop an optimized FPGA accelerator to achieve low-latency  mini-batch GNN inference (Section \ref{subsec:proprosed-approach}).

\begin{table}[]
\centering
\caption{Notations}
\vspace{-0.3cm}
\begin{adjustbox}{max width=0.48\textwidth}
\begin{tabular}{cc|cc}
\toprule
 \textbf{{Notation}} & \textbf{{Description}}  & \textbf{{Notation}}  & \textbf{{Description}} \\
 \midrule
\midrule
{$  \mathcal{G}(\mathcal{V},\mathcal{E})$ }& {input graph}  &  $ v_{i}$ & {$i^{\text{th}}$ vertex} \\ \midrule
$ \mathcal{V}$ &  {set of vertices} &  $ e_{ij}$ & {edge from $ v_{i}$ to $  v_{j}$} \\ \midrule
$ \mathcal{E}$& {set of edges} &  $ L$&{number of GNN layers} \\ \midrule
$ N$& {\# of vertices in the receptive field} &  $ \mathcal{N}_{L}(i)$& $L$-hop neighbors of $ v_{i}$ \\ \midrule
$ \bm{h}_{i}^{l}$& feature vector of $ v_{i}$
at layer $l$    & $\mathcal{N}$  & receptive field  \\  \midrule
$\bm{H}$ & vertex feature matrix & $\bm{W}^{l}$ & weight matrix of layer $l$ \\
 \bottomrule
\end{tabular}
\end{adjustbox}
\vspace{-0.3cm}
\label{tab:notations}
\end{table}

\subsection{Graph Neural Network}
\label{subsec:GNN}
 The related notations are defined in Table \ref{tab:notations}. Graph Neural Networks  (GNNs) \cite{kipf2016semi, hamilton2017inductive} are proposed for representation learning on graphs. GNNs operate on the graph $  \mathcal{G}(\mathcal{V},\mathcal{E})$ and 
 follow the  message-passing paradigm that vertices recursively aggregate information from the neighbors. 
 As shown in Algorithm \ref{algo:Aggregate-update}, $\bm{h}^{L}_{v}$ denotes the last-layer embedding of the target vertex $v$. 
%  The outputs of GNN are the embeddings of target vertices that 
 The GNN output $\bm{h}^{L}_{v}$
 can be used for many downstream tasks, such as node classification \cite{hamilton2017inductive,kipf2016semi}, link prediction  \cite{zhang2018link}, graph classification \cite{ying2018hierarchical}, etc.

\begin{small}

 \begin{algorithm}
 \caption{Recursive message-passing paradigm of GNN}
 \begin{algorithmic}[1]
 \renewcommand{\algorithmicrequire}{\textbf{Input:}}
\renewcommand{\algorithmicensure}{\textbf{Output:}}
 \Require Input graph: $\mathcal{G}(\mathcal{V},\mathcal{E})$; Initial vertex features of input graph: $\left\{\bm{h}^{0}_{1}, \bm{h}^{0}_{2}, \bm{h}^{0}_{3}, ..., \bm{h}^{0}_{|\mathcal{V}|}\right\}$; Set of target vertices: $\{v_{1}, v_{2}, ..., v_{m}\}$
 \Ensure Output embeddings: $\left\{\bm{h}^{L}_{v_{1}}, \bm{h}^{L}_{v_{2}}, \bm{h}^{L}_{v_{3}}, ..., \bm{h}^{L}_{v_{m}}\right\}$
  \For{$v_{i} \in \{v_{1}, v_{2}, ..., v_{m}\}$} 
    \For{$l =1$ to $L$ } 
        \For{$v_{j} \in \mathcal{N}_{L-l}(i)$  } 
            \State $\bm{z}_{j}^{l} = \text{aggregate}\left(\bm{h}_{k}^{l-1}:k\in  \mathcal{N}_{1}(j) \bigcap \mathcal{N}_{L-l+1}(i)\right)$
            \State $\bm{h}_{j}^{l} = \text{update}\left(\bm{z}_{j}^{l}, \bm{h}_{j}^{l-1}, \bm{W}^{l}\right)$
        \EndFor
    \EndFor
  \EndFor
  \end{algorithmic} 
 \label{algo:Aggregate-update}
 \end{algorithm}
\end{small}
 
 \vspace{0.1cm}
  In an $L$-layer  model, a target vertex $v$ aggregates information from the neighbors within $L$-hop. The neighbors within $L$-hop form the \emph{receptive field} of $v$. 
The number of vertices $N$ in the receptive field grows exponentially with the depth $L$ of the model: $N \approx \mathcal{O}(d^L)$, where $d$ is the average degree of the graph.  We denote GNNs following such recursive message-passing paradigm as \emph{Coupled models} since the size of receptive field $N$ depends on the model depth $L$. The GNN model weights includes the weight matrix $\bm{W}^{l}(1\leqslant l \leqslant L)$ of each layer and the size of weight matrices are independent of graph size.

\noindent \textbf{Specification of a Coupled model}: 
%We summarize the abstraction of a Coupled GNN model. 
A Coupled GNN model is specified by: (1) number of layers $L$, (2) $\text{aggregate}()$ function that defines operator for aggregating the neighbor information (e.g., $\text{aggregate}()$ of GCN \cite{kipf2016semi}: $\bm{z}_{j}^{l} = \text{Sum}\big(\frac{1}{ \sqrt{D(j)\cdot D(i)}} \cdot \bm{h}_{i}^{l-1}:v_{i}\in \mathcal{N}_{1}(j)\cup  \{j\} \big)$), (3) hidden dimension of each layer $f_{l}$ for $0\leqslant l \leqslant L$, (4) $\text{update}()$ function (e.g., $\bm{h}_{j}^{l}=ReLU(\bm{W}^{l}\bm{h}_{j}^{l-1})$) with weight matrix $\bm{W}^{l}~(1\leqslant l \leqslant L)$.

\begin{figure}[h]
     \centering
     \includegraphics[width=8cm]{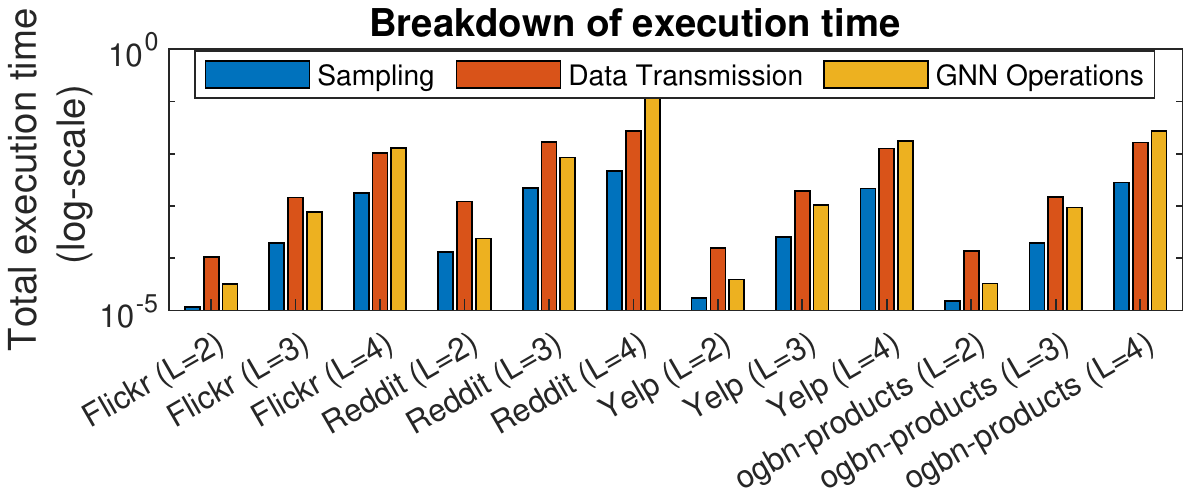} \\
     \includegraphics[width=8cm]{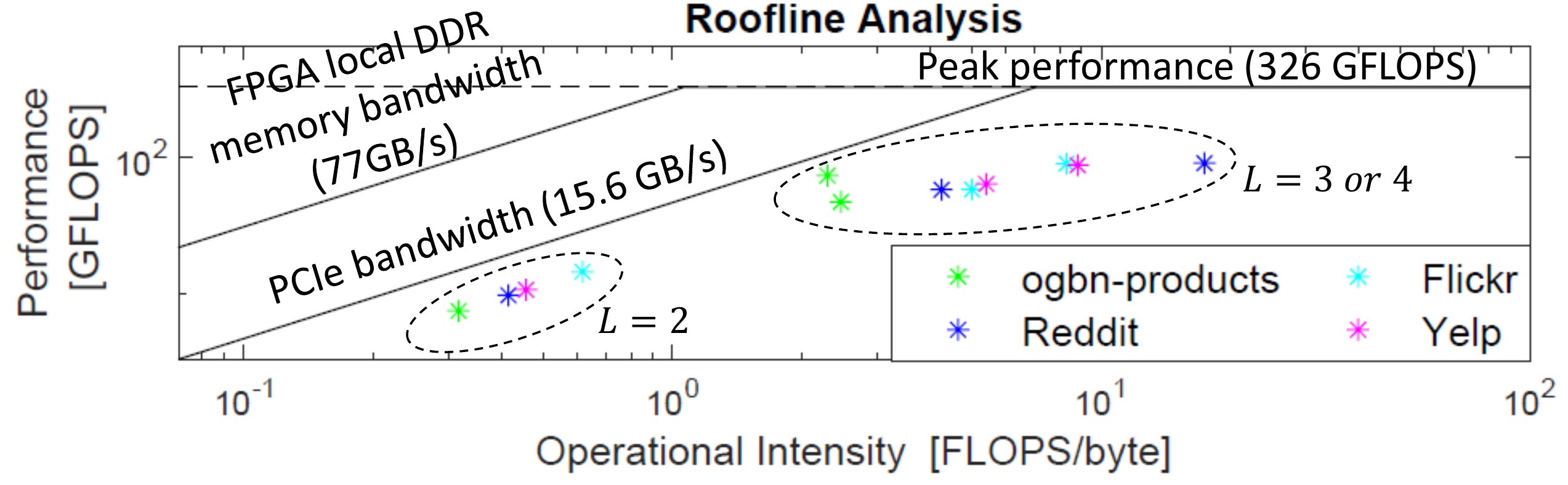} \\ 
     \vspace{-0.3cm}
     \caption{Experimental analysis of mini-batch inference using Coupled GraphSAGE \cite{kipf2016semi} model: (1) Breakdown of execution time, (2) Roofline analysis (vertical axis is in log-scale)}
     \vspace{-0.3cm}
     \label{fig:experimental-analysis}
 \end{figure}

To understand the low C2C ratio of Coupled GNNs, we profile the execution of mini-batch inference of GraphSAGE \cite{hamilton2017inductive} using a prior FPGA accelerator (GraphACT)\footnote{GraphACT is an accelerator for training the GraphSAGE model, including forward propagation for inference and back propagation for calculating weight gradients. In Figure \ref{fig:experimental-analysis}, we only perform the forward propagation of GraphACT for inference.} \cite{zeng2020graphact}  using the CPU-FPGA platform (Baseline 3) described in Section \ref{subsec:baseline}. 
The graph is stored in the external memory of the host processor, since the sizes of realistic graphs are often much larger than the on-chip capacity of GPUs and FPGAs. We further perform vertex sampling on the $L$-hop neighborhood (following the recommended parameters \cite{hamilton2017inductive}) to optimize performance.
% Four datasets are used including Reddit \cite{hamilton2017inductive}, Flickr \cite{zeng2019graphsaint}, Yelp \cite{zeng2019graphsaint}, and ogbn-products \cite{hu2020open}. 
See descriptions of the four datasets in Section \ref{tab:datasets-statistics}.
As shown in Figure \ref{fig:experimental-analysis}-(1), the data transmission between CPU and FPGA incurs significant execution time overhead because the number of neighbors grows exponentially with the depth of GNN model. 
The execution time also increases exponentially with the GNN depth. The roofline analysis (Figure \ref{fig:experimental-analysis}) demonstrates that mini-batch inference of Coupled GNN model is memory-bound, and the overall performance is limited by the available PCIe bandwidth. Moreover, the hardware accelerator has low utilization $<30\%$ (computed by $\frac{\text{achieved performance}}{\text{peak performance}}\times 100\%$).

\subsection{Model Depth-receptive Field Decoupling}
\label{subsec:decoupled-model}

%  \begin{figure}[h]
%     \centering
%     \includegraphics[width=8cm]{pic/decoupled-model.pdf}
%     \vspace{-0.3cm}
%     \caption{ An example of the Decoupled GNN models ($L$=2 and $L$=3).}
%     \vspace{-0.3cm}
%      \label{fig:decoupled model}
% \end{figure}

\begin{small}
\begin{algorithm}
 \caption{Inference process of Decoupled GNN models}
 \begin{algorithmic}[1]
 \renewcommand{\algorithmicrequire}{\textbf{Input:}}
\renewcommand{\algorithmicensure}{\textbf{Output:}}
 \Require $\mathcal{G}(\mathcal{V},\mathcal{E}, \bm{X}^{0})$; Number of layers $L$; Size of receptive field $N$; A batch of target vertices $\mathcal{V}_{t}$; GNN layer operators (aggregate(), update());
 \Ensure Node embeddings of target vertices: $\{h_{v}^{L}:v\in \mathcal{V}_{t}\}$
  \For{$v\in \mathcal{V}_{t}$ }   
    \State Identify $N$ important neighbors $\mathcal{N}_{imp}(v)$ for $v$.
    \State Build the vertex-induced subgraph $\mathcal{G}'(v)$ using $\mathcal{N}_{imp}(v)\bigcup\{v\}$
    \State Extract the input vertex features $\mathcal{F}(v) = \{h_{u}^{0}:u\in \mathcal{G}'(v)\}$.
    \For{$l\leftarrow 1$ to $L$ } 
        \State Message passing within $\mathcal{G}'(v)$ using the layer-$l$ operators
    \EndFor
    % \State Propagate the features in $\mathcal{G}'(v)$ using the GNN layer operators  for $L$ layers.
    \State Obtain the node embedding of vertex $v$ through $\text{Readout}()$. 
  \EndFor
  \end{algorithmic} 
 \label{algo:dcoupled-GNN}
 \end{algorithm}
 \end{small}

Recently, \cite{zeng2021decoupling} proposed a decoupling principle where the GNN depth $L$ and the receptive field size $N$ are specified independently. 
Decoupling is proposed based on the observation that in the Coupled GNN models, most neighbors involved in message-passing do \emph{not} provide useful information. 
Therefore, the key is to identify the important neighbors of the target vertex before applying message passing. 
As shown in Algorithm \ref{algo:dcoupled-GNN}, we first identify the important neighbors of the target $v$, denoted as $\mathcal{N}_{\text{imp}}(v)$.  Then, we build a vertex-induced subgraph $\mathcal{G}'(v)$ from $\mathcal{N}_{\text{imp}}(v)\bigcup \{v\}$. 
Next, the GNN message passing is performed within $\mathcal{G}'(v)$ for $L$ layers using the GNN layer operators. 
The node embedding $\bm{h}_{v}^{\text{emb}}$ is generated via applying the $\text{Readout}()$ function (e.g., $\text{Max}()$) to the outputs of the last GNN layer.
% The $L$-layer message passing generates the node embedding $\bm{h}_{v}^{\text{emb}}$ for the target vertex through $\text{Readout}()$ function (e.g. $\text{Max}()$, $\text{Mean}()$). 
For example, $\bm{h}_{v}^{\text{emb}}=\text{Max}(\{\bm{h}_{u}^{L}:u\in \mathcal{N}_{i}(v)\bigcup\{v\}\}$). 
Figure \ref{fig:decoupled model} shows an example. 
Note that the decoupling principle can be applied to widely used models (e.g., GCN, GraphSAGE, GIN, GAT) since it does not change the GNN layer operators (e.g., aggregate and update).  
We define the GNNs constructed by the decoupling principle as the \emph{Decoupled models}.

\noindent \textbf{Specification of Decoupled model}: %Based on the above description, 
A Decoupled model is specified by: (1) number of layers $L$, (2) number of important neighbors for the target vertex $N$ (i.e., size of the receptive field), (3) the sampling algorithm to obtain the important neighbors, (4) $\text{aggregate}()$ function, (5) hidden dimension of each layer $f_{l}$, $0\leqslant l \leqslant L$, (6) $\text{update}()$ function with weight matrix $\bm{W}^{l}$, $1\leqslant l \leqslant L$.

\noindent \textbf{Accuracy of Decoupled model}: When choosing appropriate neighbors $\mathcal{N}$ (see \cite{zeng2021decoupling}), a Decoupled model in general achieves higher accuracy than the original Coupled model. See \cite{zeng2021decoupling} for detailed theoretical and empirical evaluation.

%% file: 3_Design_Methodology.tex
\vspace{-0.2cm}
\section{PROPOSED Approach}
\label{subsec:proprosed-approach}

\subsection{Overview}
\label{subsec:overview}
% To achieve low-latency inference, we choose to accelerate Decoupled GNN models because Decoupled GNN models are more scalable than original Coupled ones (See Section \ref{subsec:analysis}). 
The objective of our hardware design is to achieve low-latency  mini-batch inference of Decoupled GNN models. We define the performance metric as \textbf{latency per batch}: given a batch of $C$ target vertices and a pre-trained Decoupled GNN, \textbf{latency} is   the time duration from receiving the $C$ target vertex indices to obtaining the vertex embeddings. 

To map Decoupled GNN models on CPU-FPGA platforms, we first identify and characterize the various computation kernels of GNNs (Section \ref{subsec:Computation Kernels}). 
% For dense computation kernels,  systolic array is an efficient architecture  \cite{cong2018polysa, wang1991systolic}. For sparse computation kernels, scatter-gather paradigm \cite{zhou2019hitgraph, zhang2021boostgcn} is  efficient for sparse computation. However, while systolic array can not deal with irregular computation pattern in sparse computation, the scatter-gather paradigm is inefficient for dense computation.
Then, we design a novel \emph{unified} architecture
named Adaptive Computation Kernel (ACK, see Section \ref{sub:hardware-modules}), capable of executing both the sparse and dense computation without any runtime reconfiguration. 
Finally, we propose a design space exploration algorithm (Section \ref{subsec:design-space-exploration}) to generate a single hardware design point for various GNN models. % -- given a \emph{ target FPGA platform} and \emph{the set of arithmetic  operations required by the set of input GNNs}, our DSE method generates a single hardware accelerator on the target FPGA platform to  support this set of GNN models. 
% To support both sparse and dense computation, we design a novel unified architecture denoted as Adaptive Computation Kernel (ACK, see Section \ref{sub:hardware-modules}). The proposed ACK can not only execute dense computation with high efficiency, but also deal with the irregular computation patterns in sparse kernels.  
% ACK is a fixed hardware design that does not require any FPGA runtime reconfiguration \cite{vipin2018fpga} to support various kernels, which usually has high overhead.  
% Moreover, 
% In contrast to 
Our design is thus advantageous compared with previous FPGA accelerators (e.g., BoostGCN \cite{zhang2021boostgcn}, Deepburning-GL \cite{liang2020deepburning} , HP-GNN \cite{lin2021hp}) which require regenerating a hardware design for each GNN model. % ,  we propose a design space exploration (DSE) method (Section  \ref{subsec:design-space-exploration}) that we can design a single hardware accelerator to support various GNN models. To be specific, given a \emph{ target FPGA platform} and \emph{a set of arithmetic  operations required by a set of GNN models}, our DSE method generates a single hardware accelerator on the target FPGA platform to  support this set of GNN models. 

\subsection{Analysis of Decoupled Models}
\label{subsec:analysis}

%  Suppose the depth of GNN model is $L$. 
 Using the $L$-layer Coupled GNN model to generate embedding for a target vertex, the information in the $L$-hop neighborhood is needed. The average number of neighbors in the receptive field is $N \approx \mathcal{O}(d^L)$, where  $N$ depends on $L$ and $d$ is the average degree of the graph. 
 In a Decoupled GNN model, since $N$ and $L$ are specified independently, it has the following benefits for hardware acceleration (To simplify the analysis, we assume $f_{i}=f$ $(i=0,1,...,L)$, and illustrate using the GraphSAGE \cite{hamilton2017inductive} model):  
\begin{itemize}
    \item  The computation cost of a Decoupled model is $\mathcal{O}(NLf^{2})$ which  grows linearly with $L$. The data communication cost is $\mathcal{O}(Nf)$ which is small since the number of neighbors $N$ is small. In a Coupled model, the computation cost is $\mathcal{O}(d^{L}f^{2})$ and the communication cost is $\mathcal{O}(d^{L}f)$. 
    \item Decoupled model ($L>1$) achieves higher computation-to-communication (C2C)  ratio, $\mathcal{O}(Lf)$, since the data communication cost is fixed, and the computation cost grows linearly with $L$. For Coupled Model, the C2C ratio is only $\mathcal{O}(f)$.
    \item Since a Decoupled GNN model has small number of neighbors (100-200 neighbors) \cite{zeng2021decoupling}, a small on-chip memory can store all the intermediate results.  Coupled GNN model ($L \geqslant 3$)  requires large data communication with external memory.
\end{itemize}

To summarize, Decoupled models achieve small computation and communication cost, high C2C ratio and require small on-chip memory, making them attractive for hardware acceleration.

% To achieve  low-latency mini-batch GNN inference, we first perform GNN model transformation based on the model depth-receptive field decoupling principle \cite{zeng2020deep} of GNN. Therefore, the Decoupled GNN models can overcome the neighbor explosion issue of Couple GNN models, and are more suitable for hardware acceleration. However, general purpose processors (e.g., CPU, GPU) \cite{yan2020hygcn, geng2020awb} have low efficiency for executing GNNs because GNNs have irregular computation pattern and memory access pattern. To address this this challenge, we propose a hardware design on CPU-FPGA heterogeneous platform for mini-batch inference. The proposed hardware design exploits the features of the FPGA to achieve low-latency inference. We design a highly optimized hardware accelerator on FPGA to efficiently execute various computation kernels of GNN. 

 \begin{figure}[h]
    \centering
    \includegraphics[width=8.5cm]{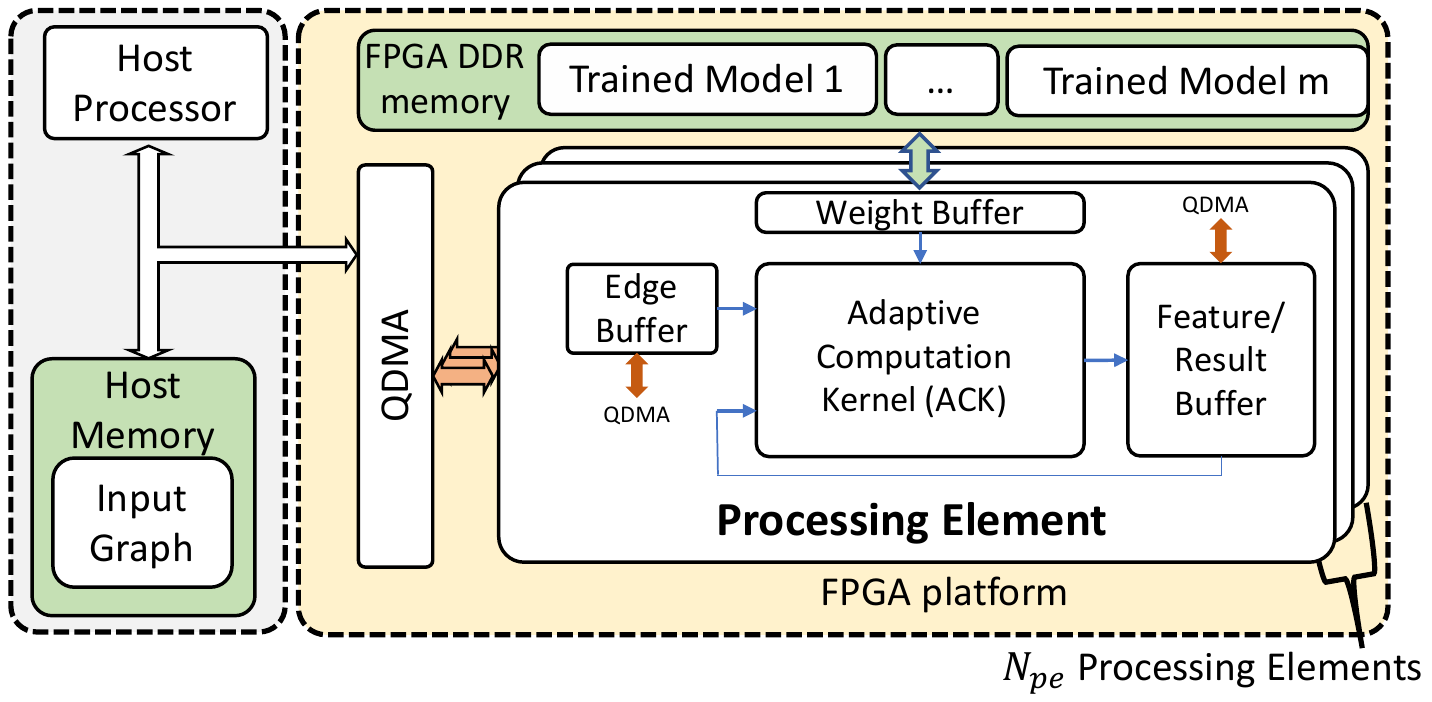}
    \vspace{-0.5cm}
    \caption{System design}
    \vspace{-0.3cm}
     \label{fig:model-architecture codesign}
\end{figure}

\noindent \textbf{Important Neighbor Identification (INI)}: INI (line 2 of Algorithm \ref{algo:dcoupled-GNN}) is the key to achieve high accuracy with a Decoupled model. 
Following \cite{zeng2021decoupling}, we use the Personalized PageRank (PPR) \cite{bahmani2010fast} score as the metric to indicate the importance of neighbor vertices w.r.t. a given target vertex. We use the local-push algorithm \cite{andersen2006local} to compute approximate PPR scores. There are several benefits of using this approach: (1) As shown in \cite{zeng2021decoupling}, PPR score is a good metric to reflect neighbor importance. Empirically, Decoupled models based on PPR  achieve high accuracy with a small number of neighbor vertices (e.g., $100-200$ vertices) \cite{zeng2021decoupling}. (2) The computation complexity of the local-push algorithm is low and remains low even when the input graph size grows \cite{aggarwal2021performance}. (3) The local-push algorithm can be easily parallelized across multiple CPU cores.

\subsection{System Design}
\label{subsec:system-design}

 Figure \ref{fig:model-architecture codesign} depicts the proposed system for executing Algorithm \ref{algo:dcoupled-GNN}.
 
 \vspace{0,1cm}
\noindent \textbf{Design Time}:
 At design time, given the specification of the target FPGA platform and a set of Decoupled GNN models (see Section \ref{subsec:design-space-exploration}), we generate a single hardware accelerator and deploy it on the target FPGA platform.  The  overhead of generating the accelerator is a one-time cost. The trained GNN models are stored in the FPGA DDR memory. User can specify which model to use at runtime. On  host platform, we develop the host program to run on the processor, which consists of three parts: (1) the subroutine for Important Neighbor Identification. (2) the subroutine  (developed using Xilinx OpenCL \cite{xilinxopencl}) for allocating tasks on FPGA accelerator for a GNN model: it takes the specification of a GNN model  and parameters (Section \ref{sec:hw-arch}) of the accelerator as input, and performs task allocation on the accelerator. A \emph{task} is a computation kernel (Section \ref{subsec:Computation Kernels}) of a GNN layer that will be executing by the accelerator. We develop a library for task allocation of the computation kernels in various GNN models. The subroutine searches for the library implementation based on the input GNN model at runtime. (3) the subroutine  to perform data communication between CPU and FPGA.

%  \begin{figure}[h]
%     \centering
%     \includegraphics[width=8cm]{pic/system-design.pdf}
%     \vspace{-0.3cm}
%     \caption{ The diagram of the proposed hardware system design.}
%     \vspace{-0.3cm}
%      \label{fig:system-deisgn}
% \end{figure}
%  \vspace{0,1cm}
% \noindent \textbf{Model Preparation}:
%  {\color{blue} Model preparation is an offline process.  To execute a specific Decoupled GNN model (in the set of Decoupled GNN models), we 
%  train this model offline and store the trained model weights in the FPGA DDR memory. We can prepare multiple models offline and store them in the FPGA DDR memory. At runtime, user can specify which model to execute from the prepared models. 
%  }
\begin{small}
\begin{algorithm}
\caption{Parallel Mini-batch Inference on CPU-FPGA}\label{alg:parallel-minibatch inference}
\begin{algorithmic}
 \renewcommand{\algorithmicrequire}{\textbf{Input:}}
\renewcommand{\algorithmicensure}{\textbf{Output:}}
\Require  A batch of target vertices $\mathcal{V}_{t}$; A Decoupled GNN model  specified by user (already trained and stored in FPGA DDR memory);
 \Ensure Node embeddings of target vertices: $\{\bm{h}_{v}^{L}:v\in \mathcal{V}_{t}\}$
\While {there is an idle CPU thread} {\color{blue}\Comment{CPU}}
    \State{Pick a target vertex $v$ from $\mathcal{V}_{t}$ and remove  $v$ from $\mathcal{V}_{t}$} 
    \State{Extract important neighbors and build vertex-induced subgraph $\mathcal{G}'(v)$}
    \State{Send  vertex features and edges of $\mathcal{G}'(v)$ to FPGA}
\EndWhile
\While {there is an idle PE}  {\color{blue}\Comment{FPGA}}
    \State{Load  vertex features and edges of $\mathcal{G}'(v)$} for a target vertex $v$
    \For{$l\leftarrow 1$ to $L$ }  {\color{blue}\Comment{Inference using ACK}}
       \For{each $kernel$ from the $kernels$ of layer $l$ }
            \State{Configure the execution mode of ACK for $kernel$}
            \State{Execute $kernel$ on ACK}
       \EndFor
    \EndFor
    % \State{Execute the inference of $v$ using the given Decoupled GNN model and obtain the embedding $h_{v}^{L}$ of $v$}
    \State{Send embedding  $\bm{h}_{v}^{L}$  back to CPU}
\EndWhile
\end{algorithmic}
\end{algorithm}
\end{small}

\noindent \textbf{Runtime}:
The overall execution process between CPU and FPGA is described in Algorithm \ref{alg:parallel-minibatch inference}. The input graph (including the edges and vertex features) is stored in the host memory. At runtime, the host processor receives the indices of a batch of target vertices and the GNN model specified by the user.
\textbf{On the host platform}, the CPU runs the host program to perform important neighbor identification (line 2 of Algorithm \ref{algo:dcoupled-GNN}) and constructs the vertex-induced subgraph for the target vertices. 
We use parallel threads on the CPU to execute  the local-push algorithm \cite{aggarwal2021performance} for multiple target vertices concurrently. Then, the CPU extracts the features of input vertices and the edges of the subgraph,  and sends them to the FPGA accelerator through the PCIe interconnection.   The CPU also performs  task allocation for the accelerator based on the specification of the GNN model. For example, for inferring a target vertex using a $L$-layer model with 2 kernels, the host program  allocates $2L$ kernels for the accelerator to execute.
% The CPU also performs task scheduling (Section \ref{sub:scheduling}) to hide the memory latency.
\textbf{On the FPGA platform}, the input data from PCIe is directly sent to the  accelerator through QDMA \cite{xilinxqdma}. The FPGA accelerator consists of $N_{\text{pe}}$ parallel and independent  processing elements (PEs),
% Each PE consists of an Adaptive Computation Kernel (ACK) to execute various computation kernels in GNNs,  an Edge Buffer to store the edges, a Weight Buffer to store weight matrices, a Feature/Result Buffer to store the vertex features. 
which can execute the forward propagation of $N_{\text{pe}}$ target vertices concurrently to obtain their embeddings. Adaptive Computation Kernel (ACK) executes the $L$-layer model layer-by-layer (see Algorithm \ref{alg:parallel-minibatch inference}). For each layer, the execution is performed kernel-by-kernel by ACK. To execute a kernel, the proper execution mode (see Section \ref{sub:hardware-modules}) of ACK is selected by setting the control bits of the hardware multiplexers in ACK. The overhead of such configuration is just one clock cycle.

%% file: 4_Hardware_Architecture.tex
\vspace{-0.1cm}
\section{Hardware Architecture}
\label{sec:hw-arch}

The proposed FPGA accelerator (Figure \ref{fig:model-architecture codesign}) consists of $N_{\text{pe}}$ parallel and independent  processing elements (PEs). Each PE has an Adaptive Computation Kernel (ACK) to execute various computation kernels in GNNs,  an Edge Buffer to store the edges, a Weight Buffer to store weight matrices, a Feature/Result Buffer to store the vertex features. The ACK of a PE contains 2-D mesh of ALUs (Section \ref{sub:hardware-modules}).
 \begin{figure}[h]
    \centering
    \includegraphics[width=8.5cm]{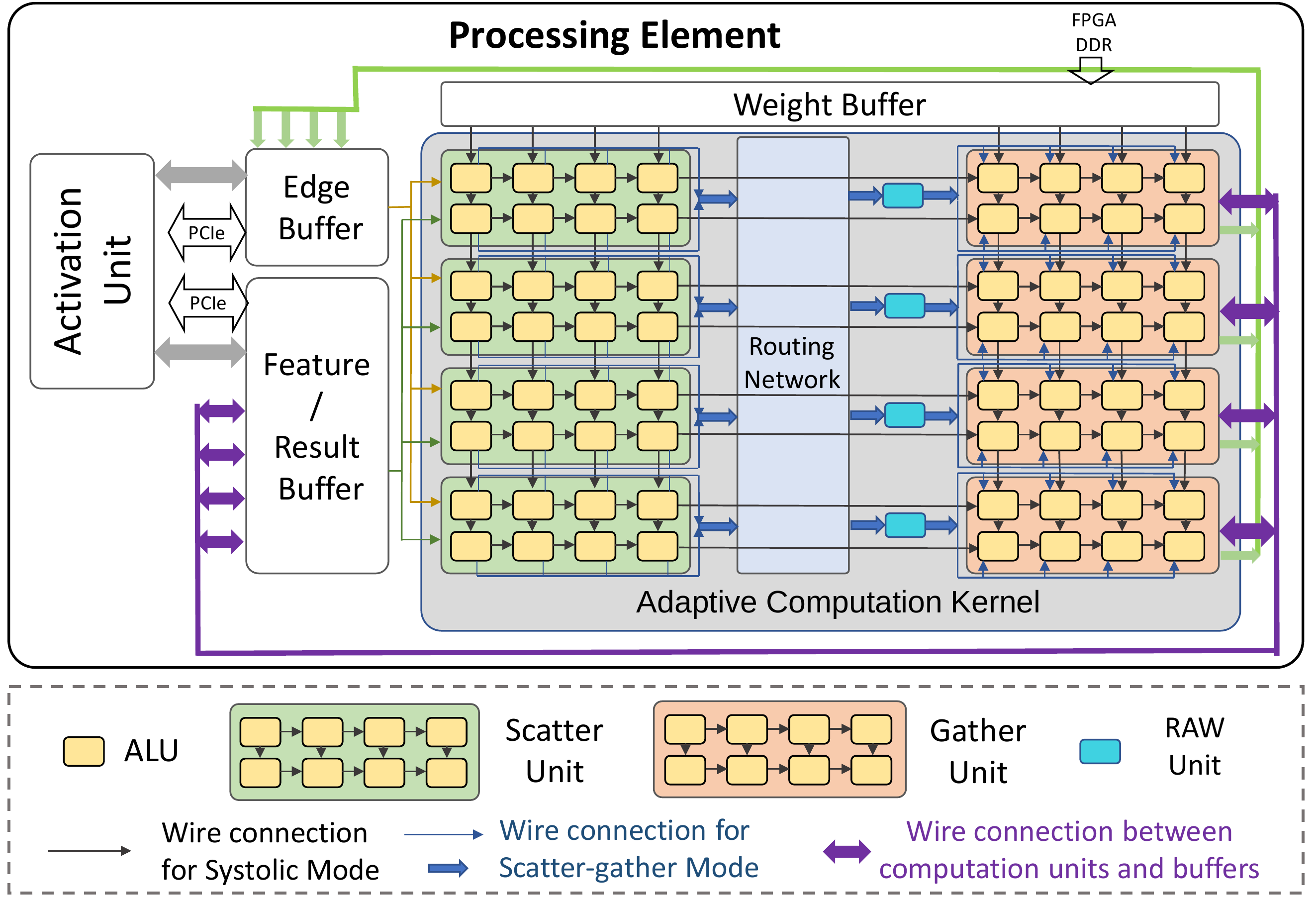}
    \vspace{-0.6cm}
    \caption{ The details of a Processing Element}
    \vspace{-0.3cm}
     \label{fig:processing-element}
\end{figure}

\subsection{Computation Kernels of GNN}
\label{subsec:Computation Kernels}
We summarize the various computation kernels in four widely used GNN models of the Decoupled version:  GCN \cite{kipf2016semi}, GraphSAGE \cite{hamilton2017inductive}, GIN \cite{xu2018powerful}, GAT \cite{velivckovic2017graph}:

\vspace{0.1cm}
\noindent \textbf{Feature Aggregation (FA)}: Feature Aggregation has two phases -- Scatter phase and Gather phase. In Scatter phase, each vertex $v_{i}$ sends its features $\bm{h}_{i}$ through its outgoing edges in the vertex-induced subgraph to its neighbors. The vertex features  are multiplied by the  edge weight to generate the intermediate results. Then, in Gather phase, each vertex aggregates the incoming intermediate results through the $\text{aggregate}()$ function (e.g., element-wise Min, Max, Mean)  to generate the aggregated features $\bm{z}_{i}$.

\vspace{0.1cm}
\noindent \textbf{Feature Transformation (FT)}: After each vertex obtaining the aggregated feature vector $\bm{z}_{i}$, the aggregated features $\bm{z}_{i}$ are transformed through the $\text{update}()$ function. In the widely used GNN models (e.g. GCN, GraphSAGE, GIN, GAT), the $\text{update}()$ is a Multi-Layer Perceptron (MLP) with an element-wise activation function (e.g., ReLU, LeakyReLU).  

\vspace{0.1cm}    
\noindent \textbf{Attention}: Some GNN models (e.g., GAT) exploit the Attention mechanism to generate data-dependent edge weights. The weight of  edge $e_{ij}$ is calculated based on  ($\bm{h}_{i}$, $\bm{h}_{j}$, $\bm{W}_{\text{att}}$, $\bm{a}$). $\bm{W}_{\text{att}}$ is the attention weight matrix that is multiplied with $\bm{h}_{i}$ and $\bm{h}_{j}$. $\bm{a}$ is the vector that is multiplied with $\bm{W}_{\text{att}}\bm{h}_{i}||\bm{W}_{\text{att}}\bm{h}_{j}$ to get the edge weight $e_{ij}$. 

\vspace{0.1cm}
While FT and Attention are dense computation kernels involving dense matrix multiplication, FA is the sparse computation kernel due to the sparsity and irregularity of the graphs. If we execute the different kernels using different hardware modules, the load imbalance can lead to hardware under-utilization  and increased latency (See Section \ref{subsec:load-balance}).

\vspace{-0.1cm}
\subsection{Hardware Modules}
\label{sub:hardware-modules}

% To support various computation kernels in GNN, previous work such as HyGCN \cite{yan2020hygcn}, GraphACT \cite{zeng2020graphact}, boostGCN \cite{zhang2021boostgcn} designed hybrid accelerators that for each kernel, a hardware module is designed and initialized independently. For example, in  GraphACT \cite{zeng2020graphact}, the Feature Aggregation Module executes FA and Feature Transformation Module executes FT. However, it is challenging to keep load balance of various computation kernels in a hybrid accelerator, because the workload of FA is unpredictable and it depends on the connectivity of the vertex-induced subgraph. If subgraph is dense, FA will have high workload and if the subraph is sparse, FA will have low
% workload. The load imbalance leads to hardware under-utilization and extra latency for mini-batch inference. 

% As discussed in Section \ref{sec:intro}, hybrid accelerators \cite{yan2020hygcn, zeng2020graphact, zhang2021boostgcn} suffer from load imbalance issue that leads to extra latency for inference. In contrast to previous works, 

To address the load imbalance challenge, we propose  Adaptive Computation Kernel (ACK) to execute various computation kernels of GNNs in a single hardware module. 

% \vspace{0.1cm}
% \noindent \textbf{Processing Element (PE)}: Our design on FPGA  (Figure \ref{fig:model-architecture codesign}) contains multiple parallel Processing Elements (PEs).  Each PE executes the message passing process of GNNs for a target vertex (Line 5 of Algorithm \ref{algo:dcoupled-GNN}). Supposing there are $N_{\text{pe}}$ PEs, the accelerator can process $N_{\text{pe}}$ target vertices concurrently. Figure \ref{fig:processing-element} depicts the architecture of a Processing Element. 

\vspace{0.1cm}
\noindent \textbf{Adaptive Computation Kernel (ACK)}: ACK contains an array of Arithmetic Logical Units (ALUs) of size $p_{\text{sys}}\times p_{\text{sys}}$. An ALU can execute various arithmetic operations including Multiplication, Addition, Multiply-Accumulation, Min, Max, etc. The proposed ACK has two execution modes -- \emph{Systolic Mode} and \emph{Scatter-Gather Mode} -- that can support FA, FT, Attention (Section \ref{subsec:Computation Kernels}).

 \begin{figure}[h]
    \centering
    \includegraphics[width=8cm]{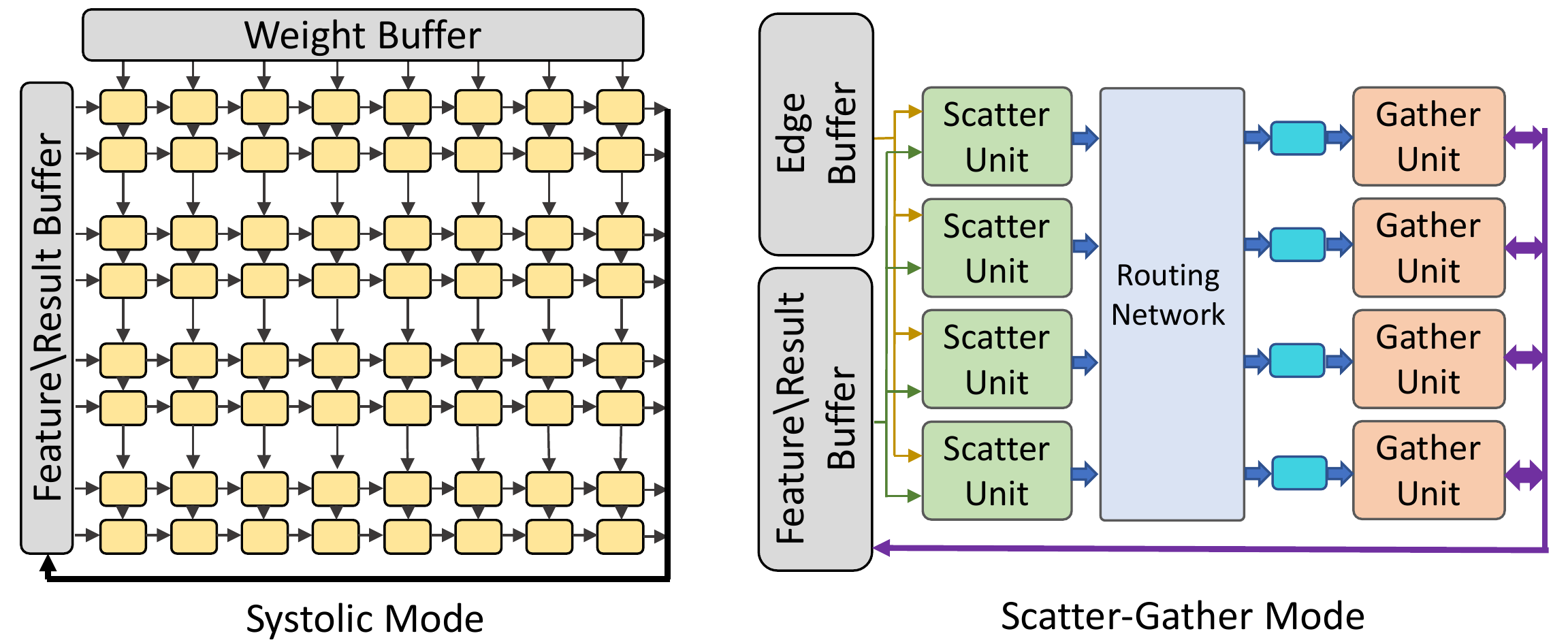}
    \vspace{-0.3cm}
    \caption{Two data paths realizing the dense (left) and sparse (right) execution modes of ACK}
    \vspace{-0.3cm}
     \label{fig:two-mode-execution}
\end{figure}

\vspace{0.1cm}
\emph{Systolic Mode}: In Systolic Mode, the array of ALUs are organized as a two-dimension systolic array.  Systolic array is an efficient architecture for dense matrix multiplication \cite{vucha2011design}, which has localized interconnections as shown in Figure \ref{fig:two-mode-execution}. Systolic Mode supports dense matrix multiplication in FT and Attention.In  Systolic Mode, ACK can execute the multiplication of weight matrix $\bm{W}$ and Feature matrix  $\bm{H}_{\text{in}}$ (See Table \ref{tab:notations}, each row of $\bm{H}_{\text{in}}$ is a vertex feature vector $\bm{h}_{i}$) to obtain the output feature matrix $\bm{H}_{\text{out}}$. Weight Buffer streams the weight matrices of MLP (FT) or the attention weight matrix to the systolic array, and Feature/Result Buffer streams multiple vertex feature vectors into the systolic array.  Systolic array of size $p_{\text{sys}}\times p_{\text{sys}}$ can execute  $p_{\text{sys}}^{2}$ Multiply-Accumulation operations in each clock cycle.  Both the Weight Buffer and the Feature Buffer have port width of $p_{\text{sys}}$ data, and can send $p_{\text{sys}}$ data to the systolic array in each clock cycle.

\vspace{0.1cm}
\emph{Scatter-Gather Mode}: In Scatter-Gather Mode, the PE executes  feature aggregation (FA) using Scatter-Gather paradigm (Algorithm \ref{alg:scatter-gather}).  The array of ALUs is partitioned into multiple Scatter Units and Gather Units. In each Scatter Unit, the ALUs are organized as a vector multiplier that multiplies the vertex feature vector by the scalar edge weight.   Similarly, in each  Gather Unit, the ALUs execute the $\text{aggregate}()$ function. Suppose the feature vector has the format $\langle src, features\rangle$, where $src$ denotes the index of the source vertex and the $features$ is the feature vector of the source vertex.  Edge has the format $\langle  src, dst, weight \rangle$, where $src, dst, weight$ denote the source vertex index, destination vertex index, edge weight respectively. The generated intermediate results (updates) by the Scatter Units have the format $\langle dst, features \rangle$.  For the vertex-induced subgraph containing $N$ vertices, the $N$ vertices are equally partitioned to the Gather Units. 
% The Routing Network routes the immediate results to the corresponding Gather Unit for feature aggregation at the destination vertex. 
The routing network performs all-to-all interconnection between Scatter Units and Gather Units. It routes the immediate results  $\langle dst, features \rangle$ generated  by Scatter Units to the corresponding Gather Units based on the index $dst$. 
For example, suppose a Gather Unit is responsible for accumulating the results to vertices $v_{1}-v_{64}$. All immediate results that has $dst$ ranging from $1$ to $64$ will be routed to this Gather Unit. The routing network is implemented as a butterfly network \cite{choi2021hbm} (See \cite{choi2021hbm} for the details of the routing network).

\begin{small}
\begin{algorithm}
\caption{Feature Aggregation using Scatter-Gather Paradigm}\label{alg:scatter-gather}
\begin{algorithmic}

\While {not done}
% \State {Scatter Unit:}
\For{each edge $e\langle src,dst,weight \rangle$ }  {\color{blue}\Comment{Scatter Unit}}
\State {Produce update $u \gets ${Scatter($src.features, e.weight$)}}
\EndFor

% \State {Gather Unit:}
\For{each update $u\langle dst,features \rangle$ }
{\color{blue}\Comment{Gather Unit}}
\State{Update vertex $dst \gets$ {Gather($u.features$)}}
\EndFor

\EndWhile
\end{algorithmic}
\end{algorithm}
\end{small}

%  \begin{figure*}[h]
%     \centering
%     \includegraphics[width=15cm]{pic/scatter-gather-example.pdf}
%     \vspace{-0.3cm}
%     \caption{ A toy example of sparse computation. }
%     \vspace{-0.3cm}
%      \label{fig:scatter-gather-example}
% \end{figure*}

%  \begin{figure*}[h]
%     \centering
%     \includegraphics[width=15cm]{pic/Task-scheduling.pdf}
%     \vspace{-0.3cm}
%     \caption{ Task scheduling for mini-batch GNN inference. }
%     \vspace{-0.3cm}
%      \label{fig:Task-Scheduling}
% \end{figure*}

% An example of the computation process is shown in Figure \ref{fig:scatter-gather-example}. There are four parallel Scatter Units and Gather Units. At each clock cycle, four edges are sent to four Scatter Units and the Scatter Units fetch the features of the source vertices. Then, four updates are generated by Scatter Units and sent to routing network. The routing network routes the updates to the Gather Units. The Gather Unit fetches the destination vertices from the Result Buffer and accumulates the updates onto the destination vertices. 
 To switch between the two execution modes, each ALU maintains multiplexers with control logic to select the input and output ports for an execution mode. When the execution of a kernel is completely finished, the ACK can start to execute the next kernel.  In our design, there are $p_{\text{sg}}$ Scatter units and $p_{\text{sg}}$ Gather Units, where $p_{\text{sg}}$ is  decided by $p_{\text{sg}} = p_{\text{sys}}/2$ . the Feature/Result Buffer have $p_{\text{sg}}$ banks. Each bank stores the feature vectors of the part of vertices in the vertex-induced subgraph (Algorithm \ref{algo:dcoupled-GNN}). Each bank is connected to a Gather Unit. Each Scatter Unit or Gather Unit has $2 p_{\text{sg}}$ ALUs.  Note that read-after-write (RAW) data hazard may occur when accumulators in the Gather Unit read the old feature vertex vector from the Feature/Result Buffer. To resolve the RAW data hazard, we implement a RAW Unit before Gather Unit. 
%  When a RAW data hazard is detected, the RAW unit will stop forwarding new update to the corresponding Gather Unit. The stalling can last for $1$ to $pip$ cycles, where $pip$ is the number of pipeline stages of a ALU.

%  \begin{figure*}[h]
%     \centering
%     \includegraphics[width=15cm]{pic/scatter-gather-example.pdf}
%     \vspace{-0.3cm}
%     \caption{ A toy example of sparse computation. }
%     \vspace{-0.3cm}
%      \label{fig:scatter-gather-example}
% \end{figure*}

 \begin{figure*}[h]
    \centering
    \includegraphics[width=15cm]{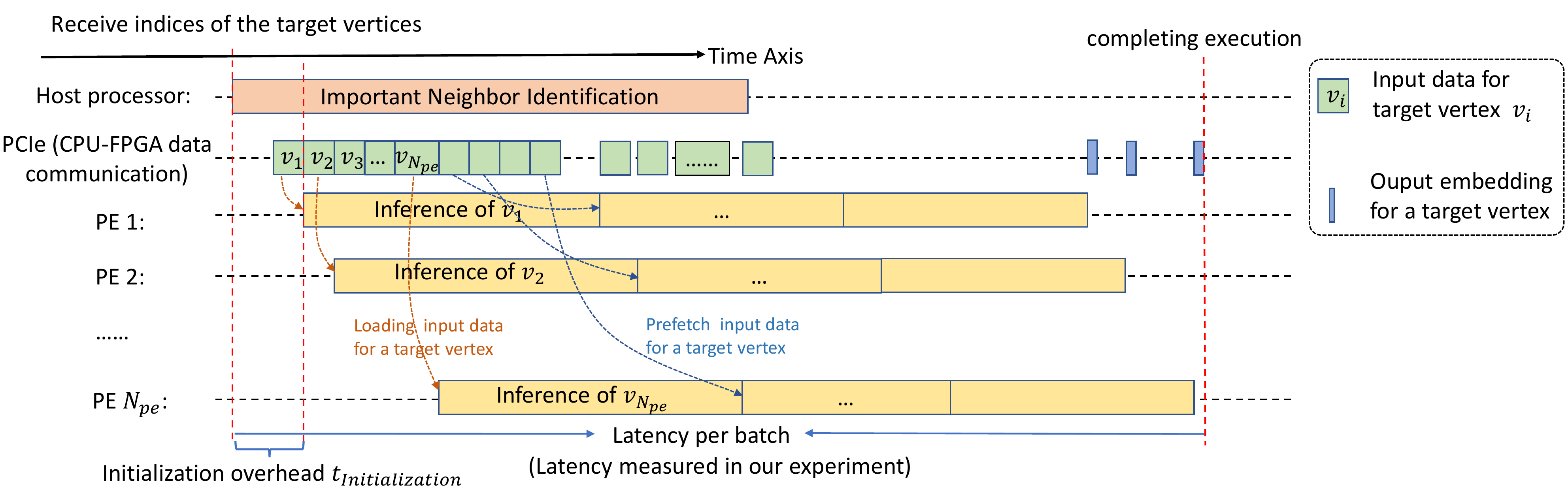}
    \vspace{-0.3cm}
    \caption{ Task scheduling for mini-batch GNN inference on the CPU-FPGA platform}
    \vspace{-0.3cm}
     \label{fig:Task-Scheduling}
\end{figure*}

\noindent \textbf{Activation Unit}: The Activation Unit executes the element-wise activation function in FT and the Softmax function in Attention. These functions are implemented using Xilinx High-Level Synthesis (HLS) \cite{o2014xilinx}. For example, the Softmax function is implemented by using $\verb|hls::exp(x)|$ function as the building block. 

\noindent \textbf{Double/triple buffering}: In a Processing Element, there are three Feature/Result Buffers for triple buffering. The first  Buffer stores the vertex feature vectors of the current GNN layer. The second Buffer  stores the  vertex feature vectors of the next GNN layer.  The third Buffer is used for prefetching the input vertex feature vectors of the next target vertex.  Similarly, Edge Buffer is also designed with triple buffering. Weight buffer is implemented using double buffering, where one buffer is used for storing the weight matrix of the current layer, and the other buffer is used to store the weight matrix of the next layer. Through double/triple buffering, memory access and computation are overlapped to reduce the overall inference latency.

\noindent \textbf{$\text{Readout}()$ function}: the $\text{Readout}()$ function (Line 6 of Algorithm \ref{algo:dcoupled-GNN}) has negligible computation complexity and it is executed by ACK in Scatter-Gather Mode. 

\subsection{Load Balance}
\label{subsec:load-balance}

The key benefit of our design is that we use a single hardware module (ACK) to execute various computation kernels with high efficiency. Therefore, we are able to assign all the on-chip  computation resources to ACKs. In the hybrid accelerators \cite{yan2020hygcn, zeng2020graphact, zhang2021boostgcn}, the  computation resources are divided among different hardware modules to execute different computation kernels. 
% Therefore, in hybrid accelerators, the load imbalance among the hardware modules can lead to hardware under-utilization.
Suppose for a single GCN layer, feature aggregatgion (FA) has the workload $\alpha_{1}$ $(\alpha_{1} >0)$ and feature transformation (FT) has the workload $\alpha_{2}$ $(\alpha_{2} >0)$, and the total computation resource is $\beta$. In our design, we use $\beta$ for ACKs. Therefore, the latency for executing this single GCN layer of our design is: $\frac{\alpha_{1} + \alpha_{2}}{\beta}$. In the hybrid accelerator, suppose the hardware module for FA uses $\beta_{1}$ resources and the hardware module for FT uses $\beta - \beta_{1}$ resources. The latency for executing this single GCN layer is $\max \left(\frac{\alpha_{1}}{\beta_{1}} , \frac{\alpha_{2}}{\beta - \beta_{1}}\right)$. It can be proved that:
\begin{equation}
    \frac{\alpha_{1} + \alpha_{2}}{\beta} \leqslant \max \left(\frac{\alpha_{1}}{\beta_{1}}, \frac{\alpha_{2}}{\beta - \beta_{1}}\right) \quad(\beta > 0, \beta_{1} >0, \beta - \beta_{1} >0)
\end{equation}
$\frac{\alpha_{1} + \alpha_{2}}{\beta} = \max \left( \frac{\alpha_{1}}{\beta_{1}}, \frac{\alpha_{2}}{\beta - \beta_{1}}\right) $
% {\color{blue}seems wrong?? $\frac{\alpha_1+\alpha_2}{\beta} = \frac{\alpha_1}{\beta}+\frac{\alpha_2}{\beta}<\frac{\alpha_1}{\beta_1}+\frac{\alpha_2}{\beta-\beta_1}$\color{black}}
when $\frac{\alpha_{1}}{\beta_{1}}  = \frac{\alpha_{2}}{\beta - \beta_{1}}$.
In the Decoupled GNN model, the workload of FA $\alpha_{1}$ is usually unpredictable because the number of edges in the receptive field depends on the connectivity of the input graph.  Moreover, varying the receptive field size can vary the workload $\alpha_{1}$, $\alpha_{2}$ at different rate. Therefore, in a fixed hybrid accelerator, it is hard to keep load balance for various input graphs and Decoupled models with various receptive field sizes. The load imbalance incurs increased latency. To execute GNN models with more than two computation kernels (e.g., GAT), load imbalance can be more severe in hybrid accelerators.

\vspace{-0.1cm}
\subsection{Task Scheduling on CPU-FPGA}
\label{sub:scheduling}

    On the CPU-FPGA platform,  the proposed task scheduling for performing inference on a batch of $C$ vertices using $N_{\text{pe}}$ parallel PEs ($C  \gg N_{\text{pe}}$) is depicted in Figure \ref{fig:Task-Scheduling}. The scheduling is based on Algorithm \ref{alg:parallel-minibatch inference}. The host processor performs Important Neighbor Identification and builds a vertex-induced subgraph for each of the target vertices. If there is an idle PE, it loads the input vertex feature vectors of the vertex-induced subgraph for a target vertex. The PE also prefetches the input data for the next target vertex. After loading the input data, the PE executes the $L$-layer GNN forward propagation for the target vertex. Finally, the PE sends the embedding of the target vertex back to the  host processor. 
    % The forward propagation of GNNs is performed layer-by-layer. 
    % Due to the proposed ACK, within each layer, various computation kernels are executed one-by-one by ACK without load imbalance among different kernels (See Section \ref{subsec:load-balance}). 

\vspace{0.1cm}
\noindent \textbf{CPU-FPGA data communication}: Using the proposed scheduling, the execution of the accelerator and the CPU-FPGA data movement are overlapped for all but the first vertex in a batch.
% There is an small initialization overhead $t_{\text{initialization}}$ for performing Important Neighbor Identification and CPU-FPGA data movement for the first vertex. 
Denote $ t_\text{initialization} = t_{\text{load}}+t_{\text{INI}}$ as the initialization overhead of a batch, where $t_{\text{INI}}$ is the latency of runing INI for a vertex using a single CPU thread on the host processor. 
Suppose that the number of important vertices for the target vertex is $N$, the feature length is $f$, each vertex feature has $b_{\text{fe}}$ bits  and each edge has $b_{\text{ed}}$ bits. The induced subgraph for a target vertex has at most $N(N-1)/2$ edges. This  data communication overhead  $t_{\text{load}}$ for loading the induced subgraph (vertex features; edges) for a target vertex can be approximated by:
\begin{equation}
 t_{\text{load}} \leq \frac{Nfb_{\text{fe}}+N(N-1) b_{\text{ed}}/2}{\text{PCIe Bandwidth}} + t_{\text{fixed}}
\end{equation}
where $t_{\text{fixed}}$ is the fixed latency for initialing a data transfer through the PCIe interconnection which is usually $0.3-0.4 \mu s$ \cite{neugebauer2018understanding}. In a Decoupled model, $N$ is small and is fixed. Thus, the above overhead is small. Note that $t_{\text{load}}$ does not increase with the depth of the GNN model. Compared with the computation complexity of the inference,  the overhead of CPU-FPGA data communication is usually negligible  (See Section \ref{subsec:breakdown-analysis}).  Thus, $t_\text{initialization}$ is negligible.

% The initialization overhead can be approximated by:
% \begin{equation}
%     t_{\text{initialization}} = t_{\text{load}}+t_{\text{INI}}
% \end{equation}
% where $t_{\text{INI}}$ is the latency of running INI for a vertex using a single CPU thread on the host processor.

\subsection{Design Space Exploration}
\label{subsec:design-space-exploration}

We perform design space exploration (DSE) to determine the hardware parameters. The inputs to our DSE are (1) available hardware resources ($N_{\text{DSP}}$: number of DSPs) on FPGA,  (2) arithmetic operations in the given set of Decoupled GNN models that needs to be supported. Given the inputs, the DSE determines the number of DSPs in a ALU $N_{\text{ALU}}$, the size of ACK in a PE $p_{\text{sys}}\times p_{\text{sys}}$, the number of PEs $N_{\text{pe}}$ in the accelerator. The proposed design has the following properties: 

\begin{itemize}
    \item  The proposed accelerator can execute a GNN model as long as the ALU can support all the arithmetic operations (e.g., Min. Max, Add, Sub, etc) in this GNN model.  $N_{\text{ALU}}$ is determined based on the arithmetic operations of a given Decoupled GNN model.
    \item The size of the ACK $p_{\text{sys}}\times p_{\text{sys}}$ in a PE determines the latency of inferring a single target vertex, and the number of PEs $N_{\text{pe}}$ decides how many target vertices can be inferred concurrently. 
    Thus, the total on-chip computation resources  should be exhausted by $N_{\text{pe}}\cdot p_{\text{sys}}^2$. 
    The value of $N_{\text{pe}}$ depends on the batch size: 
    for large batch sizes, both large and small $N_{\text{pe}}$ work well since sufficient parallelism is available across target vertices; 
    for small batch sizes, it is desirable to set $N_{\text{pe}}$ as small in order to still achieve low latency per batch. 
    Since batch sizes vary significantly in real-world applications, our DSE minimizes $N_{\text{pe}}$ by maximizing $p_{\text{sys}}\times p_{\text{sys}}$ in a PE. 
    % Given latency per batch as our performance metric, a good accelerator design should achieve low inference latency even when the batch size is small. 
    % Our performance metric is latency per batch and in real-world applications, the batch size can vary from very small to very large. For small batch size, the accelerator should still achieve low latency. To this end, we  maximize the size of ACK $p_{sys}\times p_{sys}$ in a PE and then decide the number of PEs $N_{pe}$.
    \item To efficiently implement Scatter Unit, Gather Unit and routing network in ACK on the target platform, the dimension of ACK $p_{\text{sys}}$ should be power of 2.
\end{itemize}
The above analysis leads to the following three-step DSE algorithm:
\begin{itemize}
    \item \textbf{Step 1}: Determine  $N_{\text{ALU}}$ based on all the arithmetic operations (given GNN models) to be supported.
    \item \textbf{Step 2}: Maximize the size of ALU array $p_\text{sys}$ in a PE: $p_{\text{sys}} = 2^{\left\lfloor \log_{2}\sqrt{N_{\text{DSP}}/N_{\text{ALU}}}	\right\rfloor} $
    \item \textbf{Step 3}: Determine the number of PEs: $N_{\text{pe}} = \left\lfloor\frac{N_{\text{DSP}}/N_{\text{ALU}}}{p_{\text{sys}}\times p_{\text{sys}}}\right\rfloor$
\end{itemize}
Many modern FPGAs have multiple Super Logic Regions (SLRs) with limited interconnection among SLRs. We perform the proposed three-step DSE for each Super Logic Region respectively.  The proposed DSE has constant computation complexity and can be completed using a single CPU thread instantaneously.
Note that the routing network has $p_{\text{sys}}/2$ input ports and $p_{\text{sys}}/2$ output ports with $(32\times p_{\text{sys}})$-bit data width. Its hardware cost is $\mathcal{O}( p_{\text{sys}}^2 \log p_{\text{sys}})$ which also increases with $p_{\text{sys}}$. 
% The hardware cost of routing network may potentially become the bottleneck. 
The step of maximizing $p_{\text{sys}}$ in our DSE may incur additional hardware overhead due to expanding the routing network. 
Fortunately, as shown in \cite{choi2021hbm}, even a large-scale 512-bit 32-input-32-output routing network only consumes less than 189$K$ LUTs, which is far smaller ($<18\%$) than the total LUTs of state-of-the-art FPGA boards. Since all computation is performed with ALU, as long as LUT consumption is $<100\%$ the latency won't be affected. In the large-scale FPGA device, such as Alveo U200 and U250, $p_{\text{sys}}$ does not exceed 16. Therefore, the routing network is not the resource bottleneck in our design.

% We use the High-level Synthesis (HLS) to develop the hardware templates. The obtained hardware parameters are annotated into the developed hardware templates, and we use the vendor's tool to synthesize the hardware design and generate the accelerator bitstream. Then, the bitstream  is deployed on the target FPGA platform.

\begin{figure*}[h]
     \centering
     \includegraphics[width=5cm]{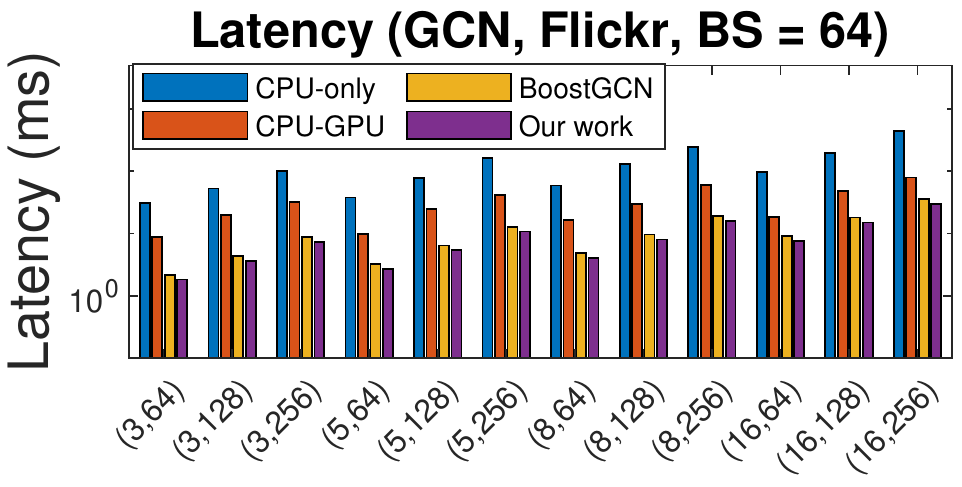} 
     \includegraphics[width=5cm]{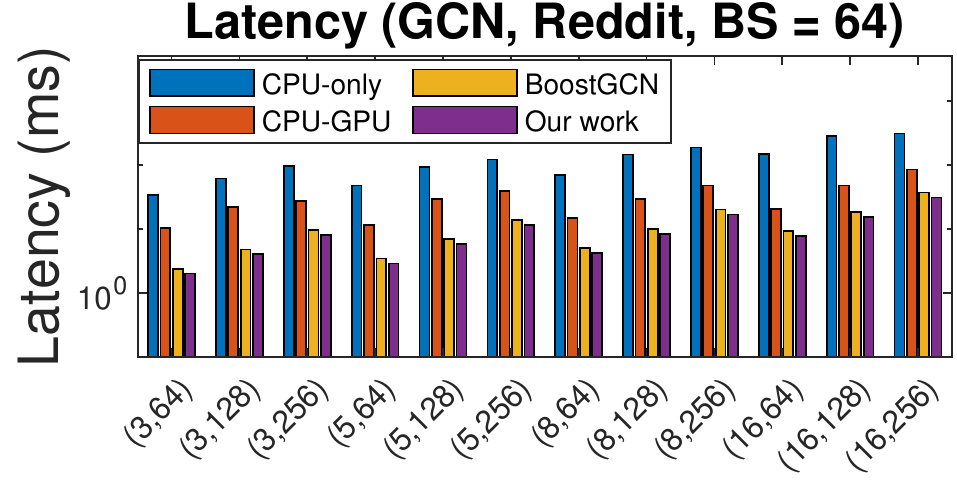} 
     \includegraphics[width=5cm]{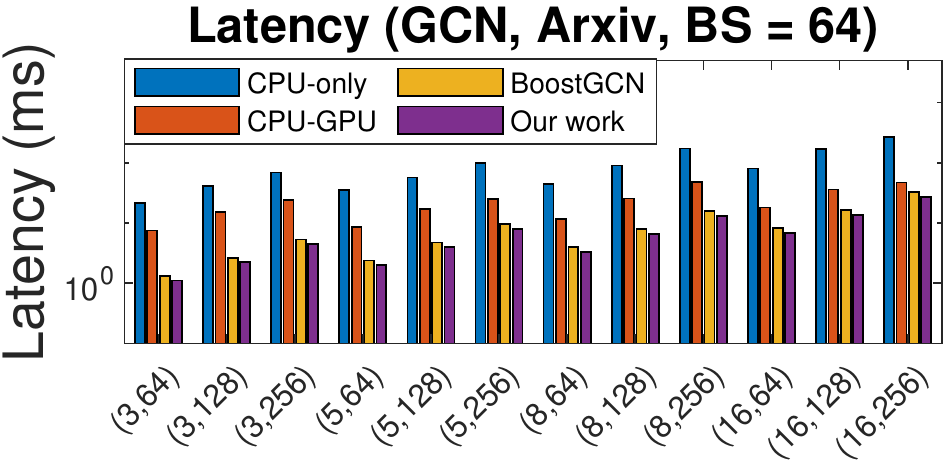} \\
      \includegraphics[width=5cm]{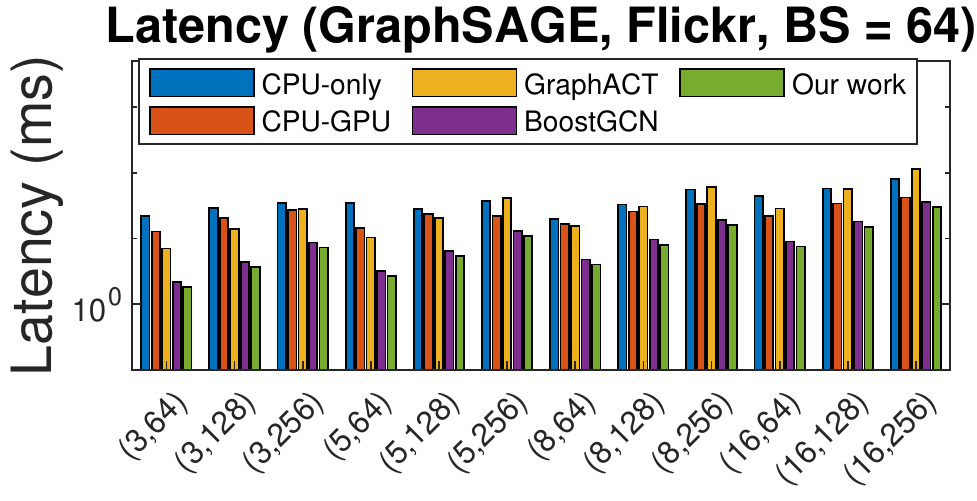} 
     \includegraphics[width=5cm]{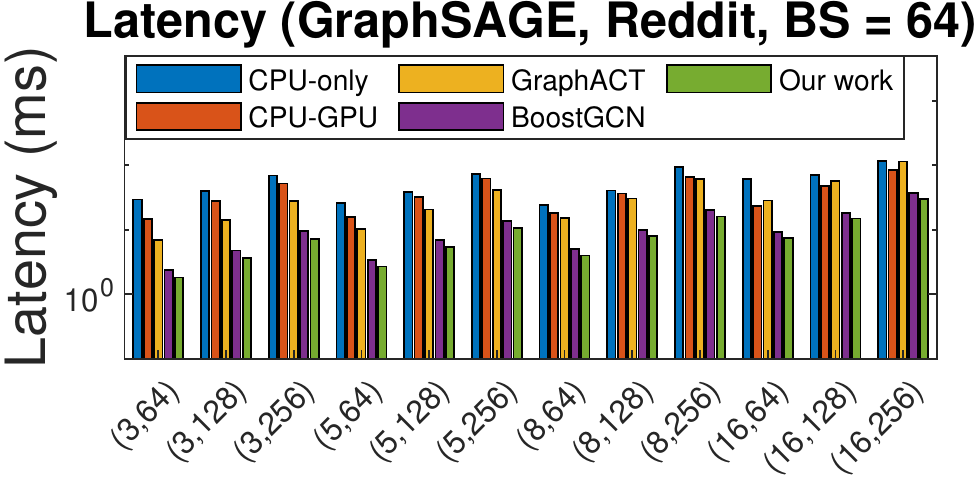} 
     \includegraphics[width=5cm]{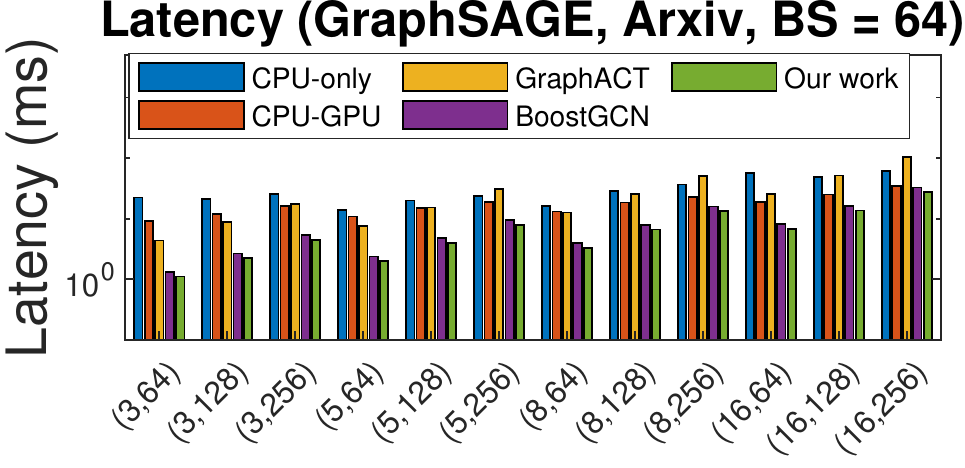} \\
      \includegraphics[width=5cm]{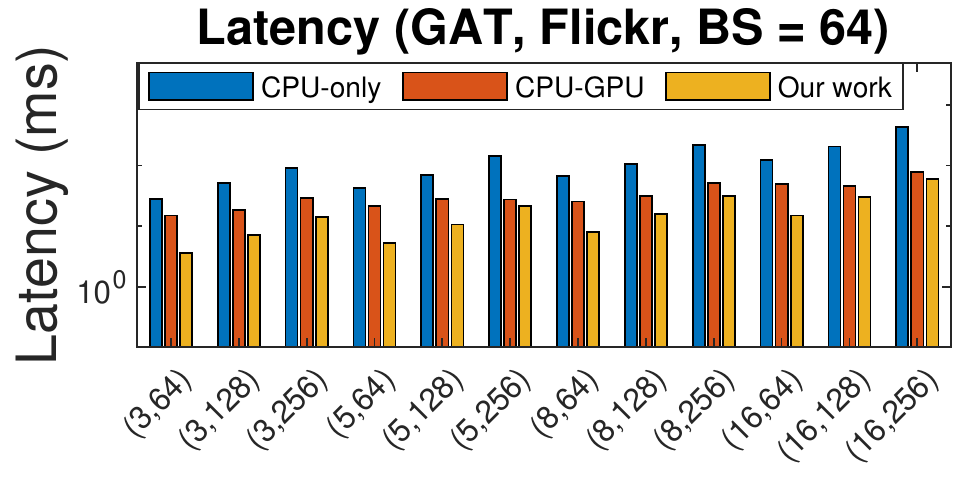} 
     \includegraphics[width=5cm]{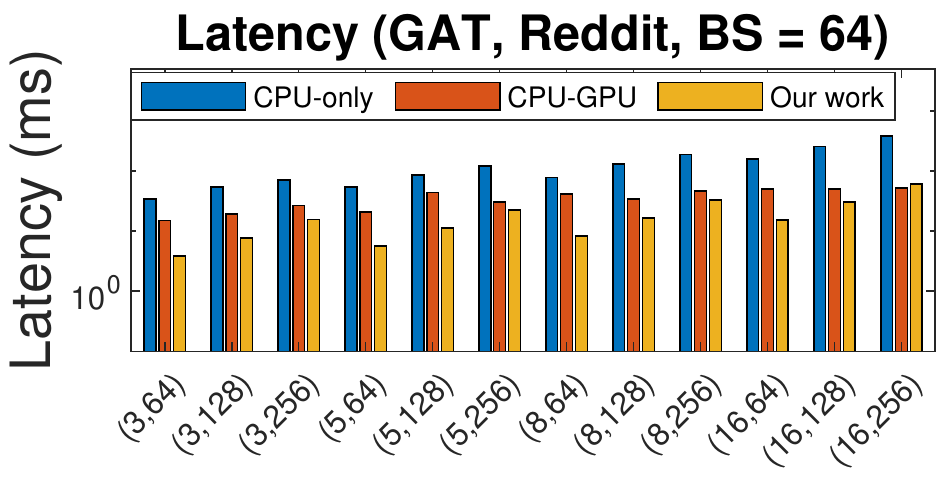} 
     \includegraphics[width=5cm]{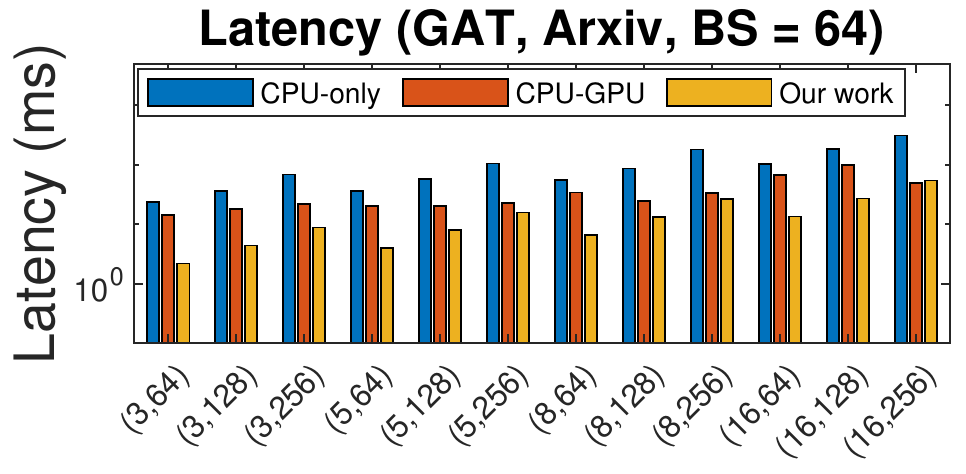} 
     \vspace{-0.3cm}
     \caption{Comparison of inference latency (Batch size=64) for  Decoupled GNN models with various depth and receptive field. Y-axis is in log-scale. X-axis denotes (number of layers $L$, size of receptive field $N$)}
     \vspace{-0.3cm}
     \label{fig:comparison results}
 \end{figure*}

%% file: 6_Experiments.tex
\section{Experiments}
\label{subsec:experiments}

\subsection{Hardware Details and Baseline Platforms}
\label{subsec:baseline}

%  \begin{figure}[h]
%     \centering
%     \includegraphics[width=6cm]{pic/two-implementations.pdf}
%     \vspace{-0.3cm}
%     \caption{ The proposed accelerator design on a state-of-the-art FPGA boards (Xilinx Alveo  U250).}
%     \vspace{-0.3cm}
%      \label{fig:accelerator-mapping}
% \end{figure}

\setlength{\columnsep}{9pt}
\begin{wrapfigure}{r}{0.25\textwidth}
  \vspace{-12pt}
%  \begin{figure}[h]
  \centering
    \includegraphics[width=4cm]{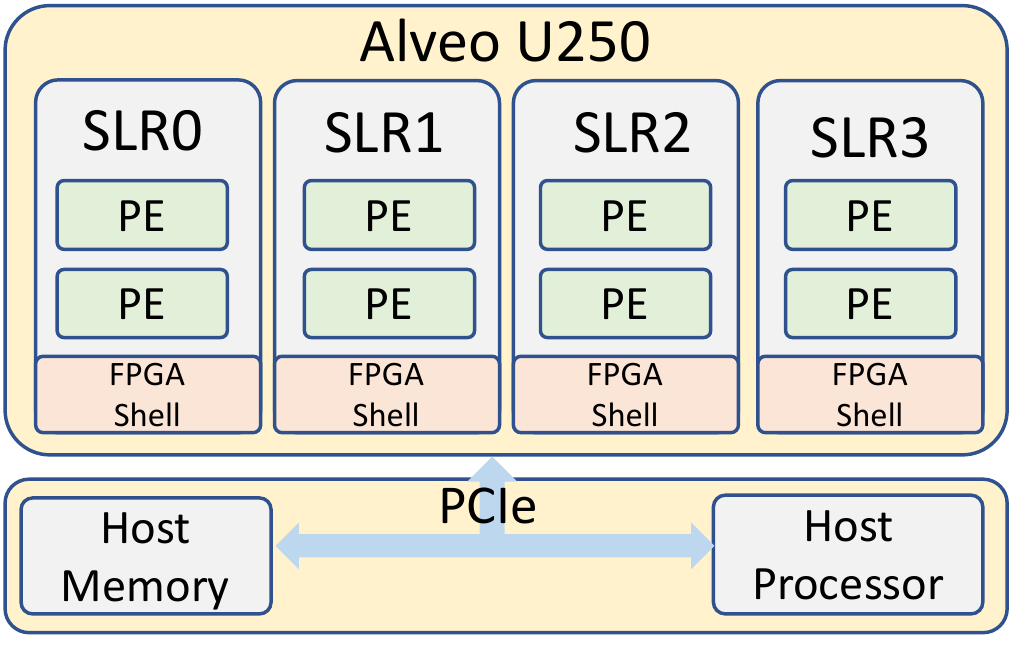}
    \vspace{-0.3cm}
    \captionof{figure}{Proposed CPU-FPGA implementation with the accelerator design on a state-of-the-art FPGA board (Xilinx Alveo  U250)}
    \vspace{-0.3cm}
     \label{fig:accelerator-mapping}
% \end{figure}
\end{wrapfigure}
We use High-level Synthesis (HLS) to develop the hardware templates. The obtained hardware parameters through DSE are annotated into the developed hardware templates, and we use the vendor's tool to synthesize the hardware design and generate the accelerator bitstream. Then, the bitstream  is deployed on the target FPGA platform.
We perform DSE to generate a hardware accelerator on a state-of-the-art FPGA platform (Xilinx Alveo U250) for three widely used GNN models (GCN, GraphSAGE, GAT). The FPGA is hosted by an Intel Xeon Gold 5120 CPU. Figure \ref{fig:accelerator-mapping} depicts the generated system design (Figure \ref{fig:model-architecture codesign}) on the CPU-FPGA platform. Based on the arithmetic operation in the three GNN models, each ALU consumes 5 DSPs. The ACK in each PE has an ALU array of size $16\times 16$.  In the ACK, there are 8 Scatter Units and 8 Gather Units. Each of the Scatter Units and Gather Units has an ALU array of size $2\times 8$. The routing network is a butterfly network of 8 input ports and 8 output ports. Each input/output port has 512-bit width. 
% On Alveo U200, there are 5 PEs in three Super Logic Regions (SLRs), since the half of the SLR1 is used for implementing FPGA Shell.
On Alveo U250, there are 8 PEs in four Super Logic Regions (SLRs) with each SLR having 2 PEs.   The hardware synthesis and Place\&Route are performed using Vitis 2021.1 \cite{xilinxvisit}. The above accelerator on Alveo U250 consumes  $762$K LUTs, $10854$ DSPs, $1853$ BRAMs and $1050$ URAMs. The resource utilization is reported after P\&R. On the host processor, we use 8 threads to execute important neighbor identification.   We deploy the host program on the host processor (Intel Xeon Gold 5120 CPU) and accelerator bitstream on the FPGA (Xilinx Alveo U250). The host processor and FPGA are connected through the PCIe $3.0\times16$ which form our target CPU-FPGA platform.  We execute the mini-batch inference on the target CPU-FPGA platform and measure the end-to-end latency using $\verb|chrono|$ library of C++.

% The accelerator on Alveo U200 consumes $747$K LUTs, $6784$ DSPs, $1211$ BRAMs and $656$ URAMs. 

% NSP = 1280 * 5 + 128*3  = 6784
%  LUT = 222000 + 259796  + 265270 = 747066

% BRAM = 1211
% URAM = 656 BRAM

% DSP = 1280 * 8 + 128 * 3 * 8 / 5 =  10854
% LUT = 481796 * 8/ 5 + 191462 = 962340

% BRAM = 1833
% URAM = 1050 BRAM

\begin{table}[!ht]
\centering
\vspace{-0.2cm}
\caption{Specifications of platforms }
\vspace{-0.3cm}
\begin{threeparttable}

\begin{adjustbox}{max width=0.47\textwidth}
\begin{tabular}{c|ccc}
 \toprule
           %\\ \midrule
\textbf{Platforms} & \begin{tabular}[|c|]{@{}c@{}} CPU \\  AMD Ryzen 3990x \end{tabular}  & \begin{tabular}[|c|]{@{}c@{}} GPU \\  Nvidia RTX3090 \end{tabular} & \begin{tabular}[|c|]{@{}c@{}} FPGA \\   Alveo U250 \end{tabular}  \\ 
\midrule \midrule 

 {Release Year}    &  2020 &  2020 & 2018   \\
 {Technology}  & TSMC 7 nm   & TSMC 7 nm & TSMC 16 nm  \\ 
{Frequency} & 2.90 GHz  & 1.7 GHz & 300 MHz 
      \\ 
{Peak Performance}& 3.7 TFLOPS & 36 TFLOPS & 0.72 TFLOPS  \\ 
{On-chip Memory}& 256 MB L3 cache & 6 MB L2 cache & 54 MB  \\
{Memory Bandwidth}& 107 GB/s & 15.6 GB/s (PCIe) & 15.6 GB/s (PCIe)\\ \bottomrule
\end{tabular}
\end{adjustbox}
\end{threeparttable}
\label{tab:platform-specifications}
\vspace{-0.3cm}
\end{table}

\noindent \textbf{Baseline Platforms}: 
We compare the following  platforms in our experiments: (1) \textbf{Baseline 1}: CPU-only platform (AMD Ryzen 3990x),
(2) \textbf{Baseline 2}: CPU-GPU platform (Intel Xeon Gold 5120 CPU + Nvidia RTX3090),
(3) \textbf{Baseline 3}: CPU-GraphACT (Intel Xeon Gold 5120 CPU + GraphACT \cite{zeng2020graphact}),
(4) \textbf{Baseline 4}: CPU-BoostGCN (Intel Xeon Gold 5120 CPU + BoostGCN \cite{zhang2021boostgcn}),
(5) \textbf{Our work}: CPU-FPGA (Intel Xeon Gold 5120 CPU  + proposed accelerator). The specifications of various platforms are shown in Table \ref{tab:platform-specifications} and Table \ref{tab:accelerators-specifications}.
% For the CPU-Only and CPU-GPU platforms, the specifications of CPU and GPU are shown in  Table \ref{tab:platform-specifications}.  The specifications of GraphACT and BoostGCN are shown in Table \ref{tab:accelerators-specifications}.
To execute mini-batch inference, the CPU-only platform uses Pytorch  with Intel MKL as the backend and the CPU-GPU plaform uses the Pytoch library with CUDA as the backend. Note that GraphACT supports GraphSAGE only. BoostGCN can support GCN and GraphSAGE. However, BoostGCN needs to generate a separate FPGA bitstream for each GNN model.

% To compare our work with GraphACT and BoostGCN, we perform following enhancements for them using our latency-reeducation strategy:
% \begin{itemize}
%     \item Add the QDMA to directly load data from the host processor to the on-chip memory of  FPGA accelerator without going through the FPGA DDR memory.
% \end{itemize}

\noindent \textbf{Latency measurement}: In the experiments, we measure the \emph{latency per batch} defined in Section \ref{subsec:overview} and Figure \ref{fig:Task-Scheduling}. For all the baselines and our work, the measured \emph{latency per batch} is the duration from the time when host processor receives the of indices of a batch of target vertices to the time when inference for all the vertices have been completed. The overhead of Important Neighbor Identification and the data movement between CPU and GPU/FPGA through PCIe are included in our measured latency.

\begin{table}[!ht]
\centering
\vspace{-0.2cm}
\caption{Platform specifications of GNN accelerators }
\vspace{-0.3cm}
\begin{threeparttable}

\begin{adjustbox}{max width=0.47\textwidth}
\begin{tabular}{c|ccc}
 \toprule
           %\\ \midrule
 & GraphACT \cite{zeng2020graphact} & This paper &  BoostGCN \cite{zhang2021boostgcn}  \\ 
\midrule \midrule 

 {Platform}    &  Xilinx Alveo U200 &  Xilinx Alveo U250 &  Intel Stratix 10 GX   \\
{Frequency} & 300 MHz  & 300 MHz & 250 MHz 
      \\ 
{Data format}& Float32& Float32 & Float32 \\
{Peak Performance}& 249.6 GFLOPS & 614 GFLOPS & 640 GFLOPS  \\ 
{On-chip Memory}& 35.8 MB  & 45 MB & 32 MB  \\
{Memory Bandwidth}& 15.6 GB/s (PCIe)  & 15.6 GB/s (PCIe) & 15.6 GB/s (PCIe) \\
 \bottomrule
\end{tabular}
\end{adjustbox}
\end{threeparttable}
\label{tab:accelerators-specifications}
\vspace{-0.3cm}
\end{table}

\begin{figure*}[h]
     \centering
     \includegraphics[width=4.2cm]{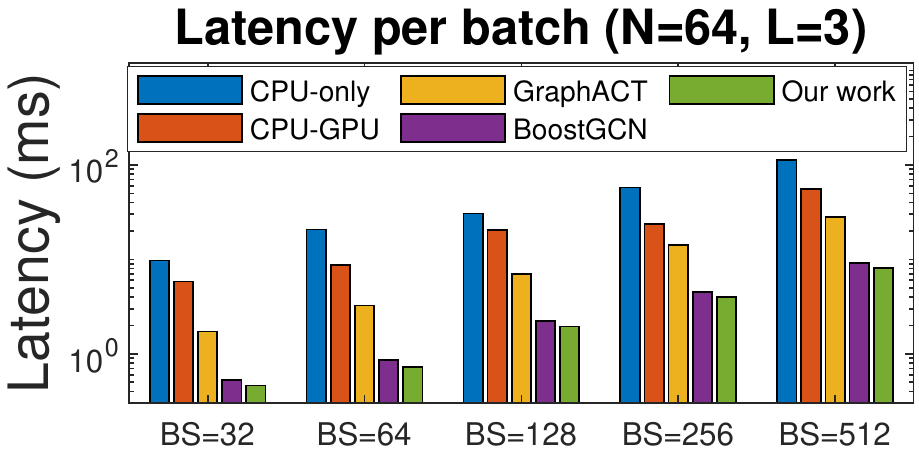} 
     \includegraphics[width=4.2cm]{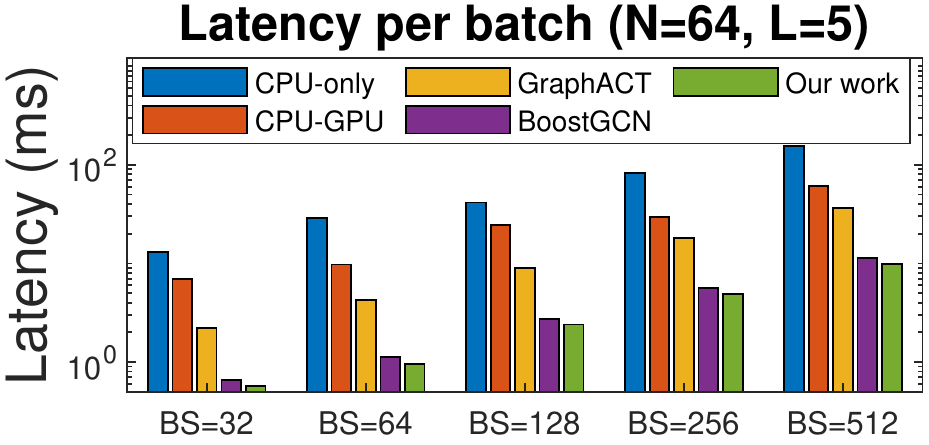} 
     \includegraphics[width=4.2cm]{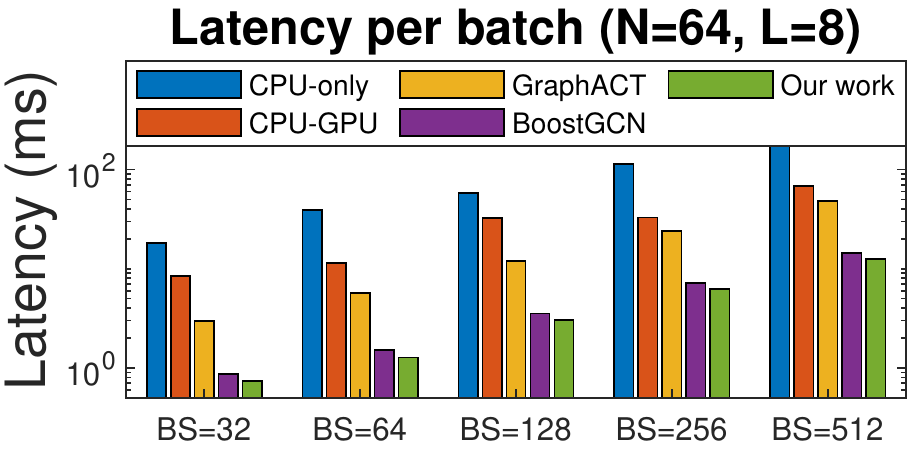} 
     \includegraphics[width=4.2cm]{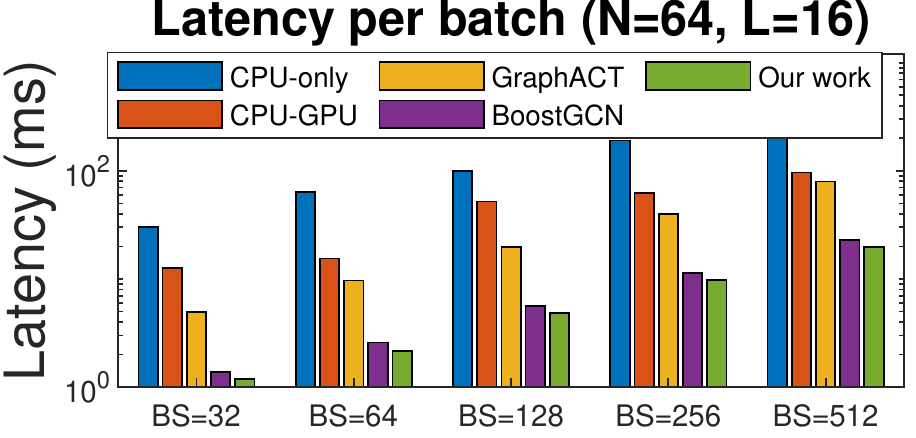} \\
     \includegraphics[width=4.2cm]{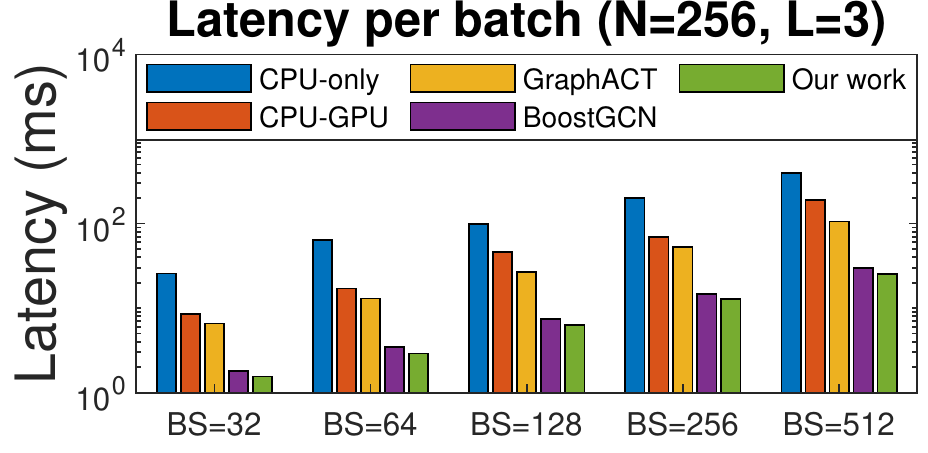} 
     \includegraphics[width=4.2cm]{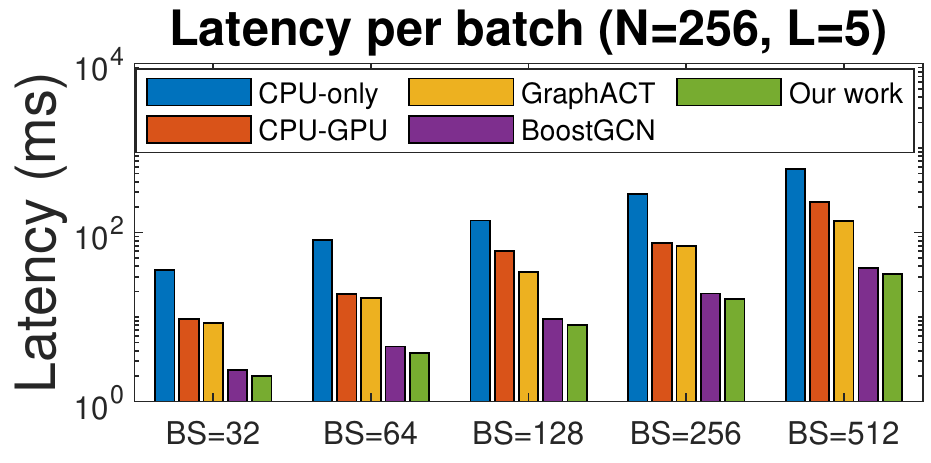} 
     \includegraphics[width=4.2cm]{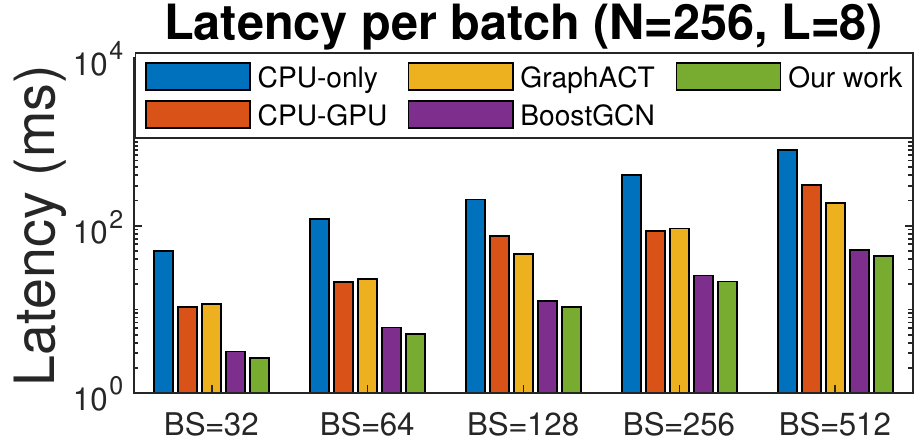}
     \includegraphics[width=4.2cm]{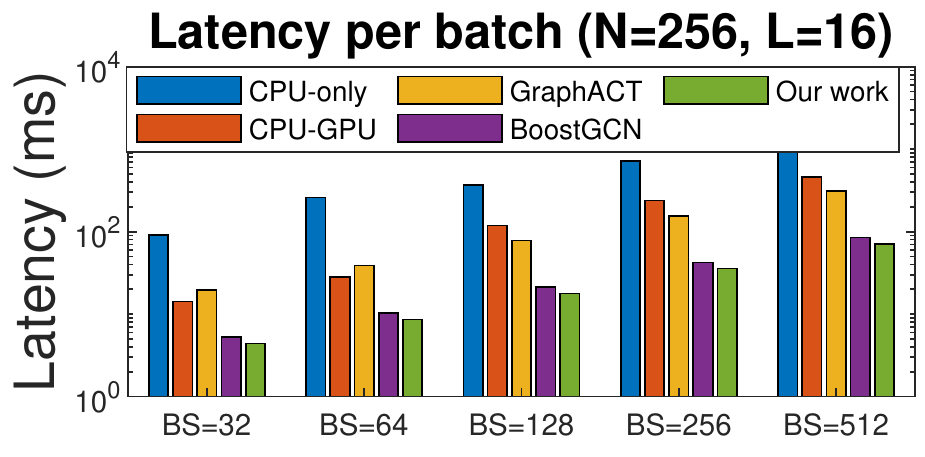} \\
     \vspace{-0.3cm}
     \caption{Latency under various batch sizes (BS) for GraphSAGE and Flickr dataset}
     \vspace{-0.3cm}
     \label{fig:various-bathsize}
 \end{figure*}

\subsection{Benchmarks}

\begin{table}[]
\centering
\vspace{-0.2cm}
\caption{Dataset Statistics}
\vspace{-0.3cm}
\begin{adjustbox}{max width=0.48\textwidth}
\begin{tabular}{cccccc}
\toprule
\textbf{Dataset}   & \textbf{Vertices} & \textbf{Edges} & \textbf{Features $f_{\text{in}}$} & \textbf{Classes} & \textbf{Degree}\\
\midrule
\midrule
Flickr (FL) \cite{zeng2019graphsaint} & 89,250 &    899,756 & 500 & 7 & 10\\
Reddit (RE) \cite{hamilton2017inductive} & 232,965  & 116,069,191 & 602 & 41 & 50                \\
ogbn-arxiv (OA) \cite{hu2020open} & 169,343 &    1,166,243 & 128 & 7 & 40 \\
\bottomrule
\end{tabular}
\end{adjustbox}
\vspace{-0.3cm}
\label{tab:datasets-statistics}
\end{table}

% \begin{table}[]
% \caption{Comparison of Accuracy ($f=256$)}
% \vspace{-0.3cm}
% \begin{adjustbox}{max width=0.48\textwidth}
% \begin{tabular}{cccccccccccc}
% \toprule
%  \multirow{2}{*}{Model}   & \multirow{2}{*}{Type}  & \multicolumn{2}{c}{Flickr }  & \multicolumn{2}{c}{Reddit} & \multicolumn{2}{c}{ogbn-arxiv}      \\ \cmidrule(lr){3-4} \cmidrule(lr){5-6} \cmidrule(lr){7-8} \cmidrule(lr){9-10} \cmidrule(lr){11-12}
%     &   & $L=3$  & $L=5$ & $L=3$  & $L=5$  & $L=3$ & $L=5$   \\ \midrule \midrule
% \multicolumn{1}{c}{\multirow{2}{*}{GCN}} & Coupled   & 0.516 & 0.522  &0.953  & 0.949  & 0.717 & 0.719 \\ \cmidrule{2-12}
% \multicolumn{1}{c}{}                     &  \begin{tabular}[|c|]{@{}c@{}} Decoupled \\ ($N=150$)\end{tabular}  & 0.549 & 0.548 & 0.957 & 0.955  & 0.728 &0.725 \\  \midrule 
% \multicolumn{1}{c}{\multirow{2}{*}{SAGE}} & Coupled & 0.514 & 0.515  & 0.965 & 0.962  & 0.719 & 0.719   \\ \cmidrule{2-12}
% \multicolumn{1}{c}{}                     &\begin{tabular}[|c|]{@{}c@{}} Decoupled \\ ($N=150$) \end{tabular} & 0.556 & 0.559  & 0.967 & 0.969  &  0.733 & 0.7333  \\
% \bottomrule
% \end{tabular}
% \end{adjustbox}
% \label{tab:comp-accuracy}
% \end{table}

We evaluate various Decoupled Models (GCN, GraphSAGE, GAT) that can achieve superior accuracy. As shown in \cite{zeng2021decoupling}, the Decoupled Models ($N<200$ and $L=3$ or $5$) can already achieve higher accuracy than the original Coupled GNN models (GCN, GraphSAGE, GAT).
% We show the accuracy of several Decoupled GNN models in Table \ref{tab:comp-accuracy}.
The Decoupled Models can achieve higher accuracy when $L$ is increased.
To evaluate Decoupled models with various $L$ and $N$, we set the hidden dimension of each GNN layer  as $f_{l}=256, (1\leqslant l \leqslant L)$ following \cite{zeng2021decoupling}. We set the number of layers $L$ as $3$, $5$, $8$, $16$ respectively.  We specify the size of the receptive field $N$ as $64$, $128$, $256$. As shown in \cite{gupta2020architectural}, recommendation systems typically use batch size $64$, $128$, $256$. 
We evaluate our design using a wider range of batch sizes $32$, $64$, $128$, $256$, $512$. We use three representative graph datasets for evaluation as listed in Table \ref{tab:datasets-statistics}. 

% We do not present the accuracy of the evaluated models. Because \cite{zeng2021decoupling} has already shown the accuracy of the Decoupled models and the purpose of this work is to accelerate the Decoupled models.

% \subsection{Computation to Communication Ratio}
% We perform the roofline analysis on the Decoupled GNN models. The results are shown in Figure \ref{fig:roofline-analysis}. Decoupled GNN models have operational intensity larger than 100 while the operational intensity of Coupled GNN models is usually small (Figure \ref{fig:experimental-analysis}). Coupled GNN models have low operational intensity because both the data communication and computation cost grows exponentially with the depth of the GNN model. For the Decoupled GNN model, the communication cost is fixed once the receptive field size is fixed.  

%  \begin{figure}[h]
%     \centering
%     \includegraphics[width=8cm]{pic/roofline-resukt.pdf}
%     \vspace{-0.3cm}
%     \caption{ Roofline analysis for the Decoupled GNN models on Xilinx Alveo U200.}
%     \vspace{-0.3cm}
%      \label{fig:roofline-analysis}
% \end{figure}

\subsection{Comparison with State-of-the-art}
We show the comparison results (latency per batch) using various GNN models, $L$ and $N$ in Figure \ref{fig:comparison results}. 
Our CPU-FPGA implementation achieves  $21.4-50.8\times$, $2.9-21.6\times$, $4.7\times$, $1.2\times$   speedup compared with CPU-only, CPU-GPU,  CPU-GraphACT, and CPU-BoostGCN,  respectively.
% After normalized using the accelerator peak performance, our CPU-FPGA implementation still achieves $1.9\times$, $1.25\times$ speedup compared with CPU-GraphACT, and CPU-BoostGCN,  respectively.
Note that BoostGCN does not support GAT and needs to generate an accelerator for each GNN model.

On the CPU-only platform, the processor can directly (without PCIe overhead) access data from the host memory  and the processor has large shared L3 cache. However, the feature aggregation of GNN results in irregular memory access pattern and low data reuse. The processor has limited L1 (32 KB) and L2 (512 KB) cache. The data exchange (vertex features;weight matrices;edges) among L1 cache, L2 cache, and L3 cache becomes the performance bottleneck and results in reduced sustained performance. For example, on multi-core platform, loading data from L3 cache incurs latency of $32$ns and loading data from L2 cache incurs latency of $5-12$ns. Compared with the CPU, the ACK in our accelerator can access data in one clock cycle during the inference execution. 

For the CPU-GPU platform, although the GPU has higher peak performance, the GPU has higher latency than our CPU-FPGA platform because: (1) GPU has extra latency of loading data from host memory to GPU global memory and loading data from GPU global memory to GPU on-chip memory, while in our CPU-FPGA implementation, the FPGA accelerator can directly load data to the on-chip memory through QDMA from the host memory. (2) Similar  to CPU, GPU has limited private L1 cache size (32 KB), therefore data exchange (vertex features;weight matrices;edges) between L2 cache and L1 cache becomes the performance bottleneck.

 We compare our CPU-FPGA implementation with GraphACT (Baseline 3) and  BoostGCN  (Baseline 4) .  GraphACT is optimized for subgraph-based mini-batch training which has similar computation pattern as the mini-batch inference of Decoupled GNN models. BoostGCN is the state-of-the-art FPGA accelerator for full-graph inference.  Compared with  CPU-GraphACT and  CPU-BoostGCN, our CPU-FPGA implementation achieves lower latency because (1) our proposed ACK can efficiently execute various kernels in GNN. GraphACT and BoostGCN follow the hybrid design that two hardware modules are initialized for feature aggregation and feature transformation, respectively. The load imbalance of the two modules leads to hardware under-utilization on GraphACT and BoostGCN. (2) We adopt the Scatter-Gather paradigm to achieve massive computation parallelism for feature aggregation. GraphACT has limited computation parallelism in its Feature Aggregation Module.

% The comparison results show that our CPU-FPGA implementation can achieve lower latency for mini-batch GNN inference than CPU-only platform, CPU-GPU platfrom and prior FPGA accelerator. 

\vspace{0.1cm}
\noindent \textbf{Latency under various batch size}: We compare the mini-batch inference latency with other platforms under various batch size. Figure \ref{fig:various-bathsize} shows the experiment results using Flickr dataset. We use the GraphSAGE model. We only show these results due to the space limitation and the observation under other experimental settings is similar. Under various batch sizes, our CPU-FPGA implementations still achieve significantly lower latency than the CPU-only platform, CPU-GPU platform, CPU-GraphACT and CPU-BoostGCN.

\subsection{Analysis of Execution Time}
\label{subsec:breakdown-analysis}

 We perform  a detailed analysis of the total execution time using the Xilinx Runtime (XRT) profiler to analyze the execution time of our   CPU-FPGA implementation. Xilinx XRT can perform fine-grained profiling for the execution time of host program, data transfer between CPU and FPGA, and the execution time of the computation kernels on the FPGA. 

\begin{figure}[h]
  \centering
    \includegraphics[width=6cm]{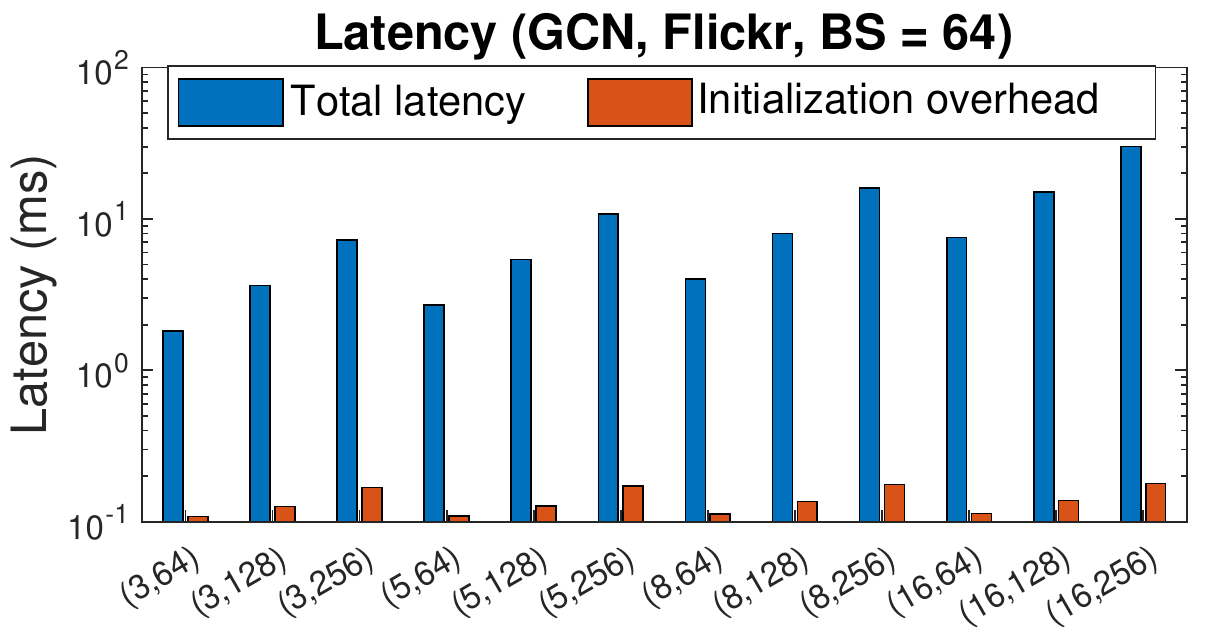}
    \vspace{-0.3cm}
    \captionof{figure}{Comparison of the initialization overhead and total inference latency under various settings}
     \label{fig:initialization-overhead}
\end{figure}

\noindent \textbf{Initialization overhead $t_{\text{initialization}}$}: We
measure the initialization overhead in our task scheduling (Figure \ref{fig:Task-Scheduling}).  We use the results in Figure \ref{fig:initialization-overhead} for illustration since other experimental settings have similar results. The initialization overhead is  $0.5\%-6\%$ of the total execution time, which is negligible.

\begin{table}[ht]
\vspace{-0.3cm}
\caption{Average latency of loading the input data for a target vertex through PCIe interconnection }
\vspace{-0.3cm}
 \begin{adjustbox}{max width=0.25\textwidth}
\begin{tabular}{cccc}
\toprule
 &  Flickr & obgn-arxiv & Reddit  \\ \midrule \midrule  
$N=64$ & $12.6$ $\mu s$ & $3.5$ $\mu s$ & $15.1$ $\mu s$  \\
$N=128$  & $29.1$ $\mu s$  & $7.7$ $\mu s$  &  $32.3$ $\mu s$ \\
$N=256$ & $72.5$ $\mu s$ & $17.1$ $\mu s$   &  $72.7$ $\mu s$\\ \bottomrule
\end{tabular}
\end{adjustbox}
\vspace{-0.3cm}
\label{tab:pcie-latency-for-data-transfor}
\end{table}

\noindent \textbf{Overhead of CPU-FPGA data communication}: To perform inference on a target vertex, the feature vectors of its  $N$ important neighbors and the edges in the induced subgraph are loaded from the host memory to the on-chip memory of a PE through  PCIe. Note that the input data are directly sent to the on-chip memory through the QDMA. As shown in Table \ref{tab:pcie-latency-for-data-transfor}, we measure the average latency for loading the input data for a target vertex through PCIe. The above latency is hidden by our task scheduling  for most target vertices (See Figure \ref{fig:Task-Scheduling}).  

 \begin{table}[ht]
  \vspace{-0.3cm}
 \caption{Overhead of INI ($t_{\text{INI}}$)}
 \vspace{-0.3cm}
 \begin{adjustbox}{max width=0.34\textwidth}
\begin{tabular}{cccc}
\toprule
 &  Flickr & ogbn-arxiv & Reddit \\
 \midrule
 \midrule
Time per vertex ($\mu s$) & $96$ & $37.6$  & $87.1$ \\
\bottomrule
\end{tabular}
\end{adjustbox}
\label{results:INI-overhead}
 \vspace{-0.3cm}
\end{table}

\noindent \textbf{Overhead of INI ($t_{\text{INI}}$)}: On  the host platform, we use 8 threads to execute INI. The measured overhead of INI $t_{\text{INI}}$ is shown in Table \ref{results:INI-overhead}. Note that the measured overhead $t_{\text{INI}}$ is the  time of INI for a vertex using single CPU thread on the host processor. The host processor can execute INI for 8 vertices concurrently. The average latency of INI is negligible compared with the total latency of  mini-batch inference ($2-100$ ms). Moreover, The overhead $t_{\text{INI}}$ for most vertices is hidden by our task scheduling (See Figure \ref{fig:Task-Scheduling}).

%% file: 7_Discussion.tex
%\section{Discussion}

%% file: 8_Conclusion.tex
\section{Conclusion}

% \color{blue}
In this paper, we proposed a novel hardware accelerator design to achieve low-latency mini-batch inference on CPU-FPGA heterogeneous platform. 
On various GNN models, we achieved 
 load-balance and high hardware utilization via the novel Adaptive Computation Kernel design.
As a result, our CPU-FPGA implementation with a single hardware accelerator (generated by our fast design space exploration algorithm) achieves significant latency reduction under various GNN models and batch sizes, compared with state-of-the-art CPU, CPU-GPU and CPU-FPGA implementations.
In the future, we plan to extend our design to a model-architecture co-design framework, such that both accuracy and latency are optimized via jointly generating the configurations of model depth, receptive field size and hardware modules. 
% \color{black}
% In this paper, we design a novel hardware accelerator on the FPGA platform for accelerating Decoupled GNN models. The proposed ACK can execute various computation kernels of GNN in a single hardware module without load imbalance. We develop a DSE algorithm to design a single accelerator to support various GNN models with FPGA reconfiguration.  The experiments show that our implementation on CPU-FPGA platform outperforms the state-of-the-art CPU-only platform, CPU-GPU platform and prior FPGA accelerators. 